\documentclass[twocolumn]{aastex62}
\usepackage{longtable}
\usepackage{lineno}
\graphicspath{{./}{figures/}}

\submitjournal{AJ}

\shorttitle{\textit{Spitzer}/IRAC Photometry of PMCs}
\shortauthors{Martinez \& Kraus}

\begin{document}

\title{A Mid-Infrared Study of Directly-Imaged Planetary-Mass Companions using Archival \textit{Spitzer}/IRAC Images}


\correspondingauthor{Raquel Martinez}
\email{raquel.martinez@utexas.edu}

\author[0000-0001-6301-896X]{Raquel A.\,Martinez}
\altaffiliation{Current address: Department of Physics and Astronomy\\
4129 Frederick Reines Hall\\
University of California, Irvine, CA 92697, USA}
\affil{The University of Texas at Austin \\
2515 Speedway, C1400 \\
Austin, TX 78712, USA}

\author[0000-0001-98111-568X]{Adam L.\,Kraus}
\affiliation{The University of Texas at Austin \\
2515 Speedway, C1400 \\
Austin, TX 78712, USA}

\begin{abstract}
The atmospheres and accretion disks of planetary-mass and substellar companions provide an unprecedented look into planet and moon formation processes, most notably the frequency and lifetime of circumplanetary disks. In our ongoing effort to leverage the extraordinary sensitivity of the \textit{Spitzer}/Infrared Array Camera (IRAC) at 3.6, 4.5, 5.8, and 8.0 $\mu$m to study wide planetary-mass and substellar companions near the diffraction limit, we present point-spread function (PSF) fitting photometry of archival \textit{Spitzer}/IRAC images for nine stars (G0 to M4+M7) in nearby star-forming regions or stellar associations that host companions at separations of $\rho = 1\farcs17-12\farcs33$. We detect all system primaries in all four IRAC channels and recover eight low-mass companions in at least one IRAC channel for our sample, five of which have not been resolved previously in IRAC images. We measure non-photospheric $[3.6]-[8.0]$ colors for four of the system companions (DH Tau B, 2M0441 B, SR 12 c, ROXs 42B b), confirming or discovering the presence of circumstellar or circum(sub)stellar disks. We detect fluxes consistent with photospheric emission for four other companions (AB Pic b, CHXR 73 b, 1RXS J1609 b, HD 203030 b) that are unlikely to host disks. Combined with past detections of accretion or disk indicators, we determine the global disk frequency of young ($<$15 Myr) wide companions with masses near the deuterium-burning limit to be $56\%\pm12\%$.
\end{abstract}


\section{Introduction} \label{sec:intro}

Dedicated exoplanet-finding surveys, such as the NASA \textit{Kepler} mission \citep{borucki10}, have revolutionized our understanding of mature planetary system populations, but the formation and evolutionary processes that lead to their properties are still not well understood. The detailed study of exoplanets on an individual basis is usually hindered by the close proximity of the exoplanet to its bright stellar host. The atmospheres of transiting exoplanets can be characterized via transmission spectroscopy \citep[e.g.,][]{deming13,kreidberg14,wakeford17}, but this limits the planetary systems that can be studied to those with semi-major axes $<$1 au. Systems with planets orbiting on wider orbits have been observed via spectroscopy behind adaptive optics (AO; \citealt{patience10,barman11a,konopacky13,haffert19,petrus20}) but these observations are difficult and expensive. Direct-imaging surveys of nearby star-forming regions have found an interesting population of wide-orbit ($>$100 au), planetary-mass companions ($<$20 $M_{\mathrm{Jup}}$; hereafter PMCs), such as 1RXS J160929.1--210524 b (8 $M_{\rm Jup}$, 330 au; \citealt{lafreniere08b}), GSC 06214--00210 B (14 $M_{\rm Jup}$, 330 au; \citealt{ireland11}), and HD 106906 b (11 $M_{\rm Jup}$, 650 au; \citealt{bailey14}), and USco1621 B (15 $M_{\rm Jup}$, 2880 au; \citealt{chinchilla20}). These systems have far more favorable separations and contrasts for the detailed study of gas giant atmospheres on larger semi-major axes. In addition, at young ages these systems also offer a unique view of moon-forming circumplanetary disks.

Most observations of directly-imaged exoplanets and planetary-mass companions have been made in the near-infrared (1--3 $\mu$m). Self-luminous exoplanets and planetary-mass objects emit substantial amounts of energy in the mid-infrared yet very few systems have been studied redward of the $L$-band (3 $\mu$m). From the ground, mid-infrared observations of exoplanets and PMCs are technically challenging, while from space, many systems were discovered after the cryogenic mission of the \textit{Spitzer Space Telescope} \citep{werner04} ended in 2009, and/or fall near or inside its diffraction limit. Extending the wavelength coverage of these objects into the mid-infrared to better fit spectral energy distributions (SEDs) will lead to more precise estimates of their physical properties and further constrain models of substellar and exoplanet atmospheres \citep[e.g.,][]{leggett08,bonnefoy10,bonnefoy14}. Utilizing the available mid-infrared observations that do exist of these systems are crucial for planning additional follow-up observations with next-generation facilities like the \textit{James Webb Space Telescope} (\textit{JWST}).

PMCs also frequently harbor disks, mostly identified through accretion signatures (e.g., line emission in H$\alpha$, Pa$\beta$, Br$\gamma$), red near-infrared colors, or mid-infrared excesses. \cite{bowler17} found $46\%\pm14\%$ of young ($<$15 Myr) substellar ($<$20 $M_{\rm Jup}$) companions with existing moderate-resolution spectroscopy had detectable Pa$\beta$ emission. This high disk frequency is comparable to that observed around isolated young substellar objects \citep{luhman10,esplin17,luhman20} but it is not clear whether wide-orbit companion disks and isolated circum(sub)stellar disks have similar accretion rates, disk compositions, and grain size distributions. Observations of PMC disks at radio wavelengths have produced only upper limits, which suggests that the dust in PMC disks might actually be more compact and optically thick \citep[e.g.,][]{bowler15,macgregor17,wolff17,wu17b}. If so, wide-orbit PMC disks are much better suited for identification and characterization in the mid-infrared.

In \cite{martinez19} (hereafter Paper I), we presented an automated point-spread function (PSF) subtraction pipeline to leverage the \textit{Spitzer} archive in the search for wide-orbit planetary-mass companions and identify excesses from circum(sub)stellar disks. Here, we apply our infrastructure to the remaining sample of known wide-orbit PMCs. In Sections \ref{spitzer_obs} and \ref{data_analysis}, we describe our sample and PSF-fitting framework. We present the results of our image analysis and pipeline performance in Section \ref{sec:results}. Finally in Section \ref{discussion}, we consider the mid-infrared photometry of the wide companions in our sample in the context of other young low-mass stars and brown dwarfs, and discuss the global disk frequency of PMCs.

\section{Sample and \textit{Spitzer} Observations} \label{spitzer_obs}
In Paper I the sample of wide-companion systems was chosen to test the feasibility of recovery via PSF-subtraction over a broad range of separations and contrast ratios. Here, we constructed a new sample to include other low-mass companions with potentially planetary mass that plausibly fit within those detection limits. We then identified systems with archival \textit{Spitzer}/Infrared Array Camera \citep[IRAC;][]{fazio04} observations from its cryogenic mission. Six of the companions have not been resolved in \textit{Spitzer}/IRAC, while three companions have had IRAC photometry reported in the literature previously. Seven of the systems belong to the young star-forming regions or stellar associations of Taurus, Carina, Chameleon, Upper Scorpius, and $\rho$ Ophiuchus, while two are young field objects. We target the young field objects because their lower distances provide good sensitivity to both mass and projected separation.

IRAC operated with four filters in the mid-infrared: 3.6, 4.5, 5.8, and 8.0 $\mu$m. The IRAC detector has 256$\times$256 pixels with a pixel scale of $1\farcs22$. We work with IRAC's cryogenic-phase corrected basic data (CBCD) and uncertainty (CBUNC) files. All data were reduced with the \textit{Spitzer} Science Center software pipeline version S18.25.0. We used the high-precision astrometry measurements of the companions from previous high-contrast AO observations as priors in our Markov Chain Monte Carlo (MCMC) fits (see Section \ref{data_analysis}).  

The combined sample primary properties are given in Table \ref{prim_tab} and system properties in Table \ref{comp_tab}. The specific details about the \textit{Spitzer}/IRAC programs and data products are listed in Table \ref{files_tab}.

\begin{deluxetable*}{llrrrccccc}[t!]
\tablecaption{Primary Properties of Directly-Imaged Substellar Companion Systems}
\tablehead{\colhead{2MASS} & \colhead{Other Name} & \colhead{$K_s$\tablenotemark{a}} & \colhead{$W1$\tablenotemark{b}} & \colhead{$W2$\tablenotemark{b}} & \colhead{SpT} & \colhead{$A_V$} & \colhead{Distance} & \colhead{Age} & \colhead{Ref.} \\
& & \colhead{(mag)} & \colhead{(mag)} & \colhead{(mag)} & & \colhead{(mag)} & \colhead{(pc)} & \colhead{(Myr)} & }
\startdata
J04294155+2632582	& DH Tau            & 8.18  & 7.65	& 7.12	& M1        & 1.4	                & $133.3\pm0.4$	         & $2\pm1$	 & 1\\
J04414565+2301580	& 2M0441 Aab   & 9.85  & 9.70	& 9.47	& M4.3; M7	& 0.2	                & $122.9^{+1.1}_{-0.9}$  & $2\pm1$	 & 2--4\\
J06191291--5803156	& AB Pic            & 6.98  & 7.28	& 6.91	& K1	    & 0.27	                & $50.13\pm0.03$         & $45\pm4$	 & 5, 6\\
J11062877--7737331	& CHXR 73           & 10.70 & 10.52	& 10.21	& M3	    & 6.5\tablenotemark{c}	& $190.0^{+3.3}_{-3.0}$  & $2\pm1$	 & 7\\
J13164653+0925269	& GJ 504            & 4.03  & 4.20	& 5.30	& G0	    & 0.0                   & $17.57\pm0.04$         & $100-6500$& 8\\
J16093030--2104589	& 1RXS J1609  & 8.92	& 8.79	& 8.78	& M0	    & 0.9	                & $137.8^{+0.3}_{-0.4}$	 & $11\pm2$	 & 9\\
J16271951--2441403	& SR 12 AB          & 8.41	& 8.29	& 8.16	& K4; M2.5  & 1.8\tablenotemark{c}	& $112.5^{+5.8}_{-5.3}$	 & $3\pm2$	 & 10, 11\\
J16311501--2432436	& ROXs 42B          & 8.67	& 8.48	& 8.37	& M0	    & 2.4	                & $145.4^{+0.5}_{-0.7}$  & $3\pm2$	 & 12, 13\\
J21185820+2613500	& HD 203030         & 6.65	& 6.98	& 6.70	& K0	    & 0.03	                & $39.23^{+0.03}_{-0.04}$& $130-400$ & 14, 15		
\enddata
\tablenotetext{a}{2MASS Point Source Catalog \citep{cutri03}}
\tablenotetext{b}{CatWISE Source Catalog \citep{marocco20}. The $W1$ and $W2$ values reported for AB Pic and GJ 504 are lower than in AllWISE \citep{cutri14} likely because of saturation.}
\tablenotetext{c}{$A_V$ converted from $A_J$.}
\tablecomments{\textit{Gaia} EDR3 parallactic distances are used from \cite{bailer-jones21} except for SR 12 AB, where we use its \textit{Gaia} DR2 parallactic distance from \cite{bailer-jones18}.}
\tablerefs{(1) \cite{itoh05};
(2) \cite{todorov10}; (3) \cite{todorov14}; (4) \cite{bowler15};
(5) \cite{chauvin05}; (6) \cite{bonnefoy10};
(7) \cite{luhman06c};
(8) \cite{kuzuhara13};
(9) \cite{lafreniere08b};
(10) \cite{kuzuhara11}; (11) \cite{bowler14};
(12) \cite{kraus14}; (13) \cite{currie14};
(14) \cite{metchev06}; (15) \cite{miles-paez17}}
\label{prim_tab}
\end{deluxetable*}

\begin{deluxetable*}{llccccc}
\tablecaption{Properties of Directly-Imaged Substellar Companions}
\tablehead{\colhead{2MASS} & \colhead{Other Name} & \colhead{Separation} & \colhead{Position Angle} & \colhead{Filter} & \colhead{$\Delta m$} & \colhead{Ref.} \\
\colhead{(Primary)}& \colhead{(Companion)} & \colhead{(arcsec)} & \colhead{(deg)} & & \colhead{(mag)} & }
\startdata
J04294155+2632582	& DH Tau B		    & $2.31\pm0.02$	    & $138.5\pm0.1$     & $K^{\prime}$  & 5.92	& 1--4  \\
J04414565+2301580	& 2M0441 Bab        & $12.325\pm0.007$	& $238.0\pm0.1$	    & $K_s$         & 3.31	& 5--7  \\
J06191291--5803156	& AB Pic b			& $5.453\pm0.025$	& $175.25\pm0.34$	& $K_s$         & 7.16  & 8, 9  \\
J11062877--7737331	& CHXR 73 b			& $1.30\pm0.03$	    & $234.9\pm1.0$	    & $K_s$         & 4.70  & 10    \\
J13164653+0925269	& GJ 504 B		    & $2.483\pm0.015$	& $326.46\pm0.36$	& $L^{\prime}$  & 12.90 & 11, 12\\
J16093030--2104589	& 1RXS J1609 b      & $2.219\pm0.002$	& $27.7\pm0.1$	    & $K_s$         & 7.25  & 13, 14\\
J16271951--2441403	& SR 12	C		    & $8.673\pm0.153$	& $166\pm2$\tablenotemark{a}& $K_s$ & 6.16	& 15, 16\\
J16311501--2432436	& ROXs 42B b		& $1.172\pm0.002$	& $270.09\pm0.17$	& $K_s$         & 6.34	& 17--19\\   
J21185820+2613500	& HD 203030	B		& $11.923\pm0.021$	& $108.76\pm0.12$	& $K_s$         & 9.56 	& 20, 21
\enddata
\tablenotetext{a}{P.A.~estimated from Fig. 1 of \cite{kuzuhara11}.}
\tablecomments{Uncertainties listed were used as input errors on the P.A.~prior.}
\tablerefs{(1) \cite{itoh05}; (2) \cite{bonnefoy14}; (3) \cite{zhou14}; (4) \cite{kraus14};
(5) \cite{todorov14}; (6) \cite{bowler15}; (7) \cite{kraus09b}; 
(8) \cite{chauvin05}; (9) \cite{bonnefoy14};
(10) \cite{luhman06c};
(11) \cite{kuzuhara13}; (12) \cite{skemer16};
(13) \cite{lafreniere08b}; (14) \cite{wu15};
(15) \cite{kuzuhara11}; (16) \cite{bowler14};
(17) \cite{kraus14}; (18) \cite{currie14}; (19) \cite{bowler14};
(20) \cite{metchev06}; (21) \cite{miles-paez17}}
\label{comp_tab}
\end{deluxetable*}

\begin{deluxetable*}{lcccccclrl}
\tablecaption{\textit{Spitzer}/IRAC Observations}
\tablehead{\colhead{2MASS} & \multicolumn{4}{c}{No.\ of Frames} & \colhead{$T_{\mathrm{exp}}$} & \colhead{AOR} & \colhead{Date}	& \colhead{PID}	& \colhead{PI} \\ \cline{2-5}
				& Ch 1	& Ch 2	& Ch 3	& Ch 4	& \colhead{(s)}					&		& \colhead{(UT)}	&			&				}
\startdata
J04294155+2632582   & 3     & 3     & 3     & 3	    & 0.4/10.4  & 3963392   & 2004 Mar 7    & 37	& G. Fazio	    \\
                    & 1     & 1     & 1     & 1     & 0.4/10.4  & 11232256  & 2005 Feb 23   & 3584  & D. Padgett    \\
                    & 1     & 1     & 1     & 1     & 0.4/10.4  & 11236096  & 2005 Feb 24   & 3584  & D. Padgett    \\
J04414565+2301580	& 0/5   & 1/5   & 0/5   & 1/5   & 1.0/26.8  & 18364160	& 2007 Mar 28	& 30540	& J. Houck      \\
J06191291--5803156	& 9	    & 9	    & 9	    & 9	    & 0.4/10.4	& 15174656	& 2005 Sep 18	& 20795	& P. Lowrance   \\
J11062877--7737331	& 2	    & 2	    & 2	    & 2	    & 0.4/10.4	& 3960320	& 2004 Jun 10	& 37	& G. Fazio  	\\
J13164653+0925269	& 1/5   & 0/5	& 1/5   & 0/5   & 1.0/26.8  & 3921920	& 2004 Jan 9	& 34	& G. Fazio      \\
                    &       & 1/5   & 	    & 1/5   & 1.0/26.8  & 18010368	& 2006 Jul 9	& 30298	& K. Luhman     \\
J16093030--2104589	& 1/5   & 	    & 1/5   & 	    & 0.4/10.4  & 15844608	& 2005 Aug 24	& 20103	& L. Hillenbrand\\
	                & 	    & 9	    & 	    & 9	    & 1.2       & 13872384	& 2006 Mar 26	& 20069	& J. Carpenter	\\
J16271951--2441403	& 2	    & 3	    & 2	    & 3	    & 0.4/10.4	& 3652096	& 2004 Mar 7	& 6     & G. Fazio      \\
                    & 2	    & 2	    & 2	    & 2	    & 0.4/10.4	& 5771008	& 2004 Mar 28	& 177	& N. Evans  	\\
J16311501--2432436	& 2/2   & 2/2   & 2/2   & 2/2   & 0.4/10.4	& 5752320	& 2004 Mar 28	& 177	& N. Evans      \\
                    & 2     & 2     & 2     & 2     & 10.4      & 5756928	& 2004 Mar 29	& 177	& N. Evans      \\
J21185820+2613500	& 16/16 & 16/16 & 16/16 & 16/16 & 1.0/26.8  & 23036416	& 2008 Jun 19	& 40489	& S. Metchev	\\
                    & 16/16 & 16/16 & 16/16 & 16/16 & 1.0/26.8	& 23796480	& 2007 Nov 15	& 40489	& S. Metchev	
\enddata
\tablecomments{}
\label{files_tab}
\end{deluxetable*}

\section{Data Analysis}
\label{data_analysis}
Previous analyses of \textit{Spitzer}/IRAC images have searched for wide-orbit PMC systems by taking advantage of IRAC's well-behaved PSF wings at $>>\lambda/D$ \citep[e.g.,][]{janson15,durkan16,baron18}. Our framework is optimized for probing the IRAC PSF at 1--5 $\lambda$/$D$, where companion identification is difficult because the PSF is undersampled at the native $1\farcs22$ pixel scale. Classical PSF-modeling techniques, such as ``locally optimized combination of images" (LOCI; \citealt{lafreniere07}) or principal component analysis, require more pixels to adequately model the primary star PSF. We use the framework described in Paper I to model the point spread functions of the system components in the IRAC images. To summarize, we use the point response function (PRF, or effective PSF; \citealt{hoffman05}) developed by the \textit{Spitzer} Science team to generate model PSFs at any position on the IRAC detector. We then fit a two-source PSF model in each image performing a MCMC analysis using a Metropolis-Hastings algorithm with Gibbs sampling. The PSF model is described by seven parameters: $x$-pixel coordinate of the primary centroid ($x$), $y$-pixel coordinate of the primary centroid ($y$), image background ($b$), peak pixel value of the primary ($n$), projected separation ($\rho$), position angle (PA), and contrast ($\Delta m$). In addition, image pixel values greater than 90\% of the saturation limit were masked. We adopt priors on separation and position angle from past high-resolution imaging results, listed in Table \ref{comp_tab}.

The MCMC analysis is conducted in two stages to determine image-specific parameters ($x$, $y$, $b$) separately from system-specific parameters ($n$, $\rho$, PA, $\Delta m$). We ran four MCMC chains with 140,000 steps each, discarding the first 10\% of each chain as ``burn-in". The weighted average median ($x$, $y$)-centroid, $\rho$, PA, and $\Delta m$ generated by the MCMC fit is used to create individual PSF models of each system component from which aperture photometry using a $10\arcsec$ radius is measured. The zero-points of IRAC Channels 1--4 are $280.9\pm4.1$, $179.7\pm2.6$, $115.0\pm1.7$, and $64.9\pm0.9$ Jy, respectively.

\begin{figure*}
    \centering
    \includegraphics[trim={0.97in 0.35in 0.75in 0.35in},clip,width=\textwidth]{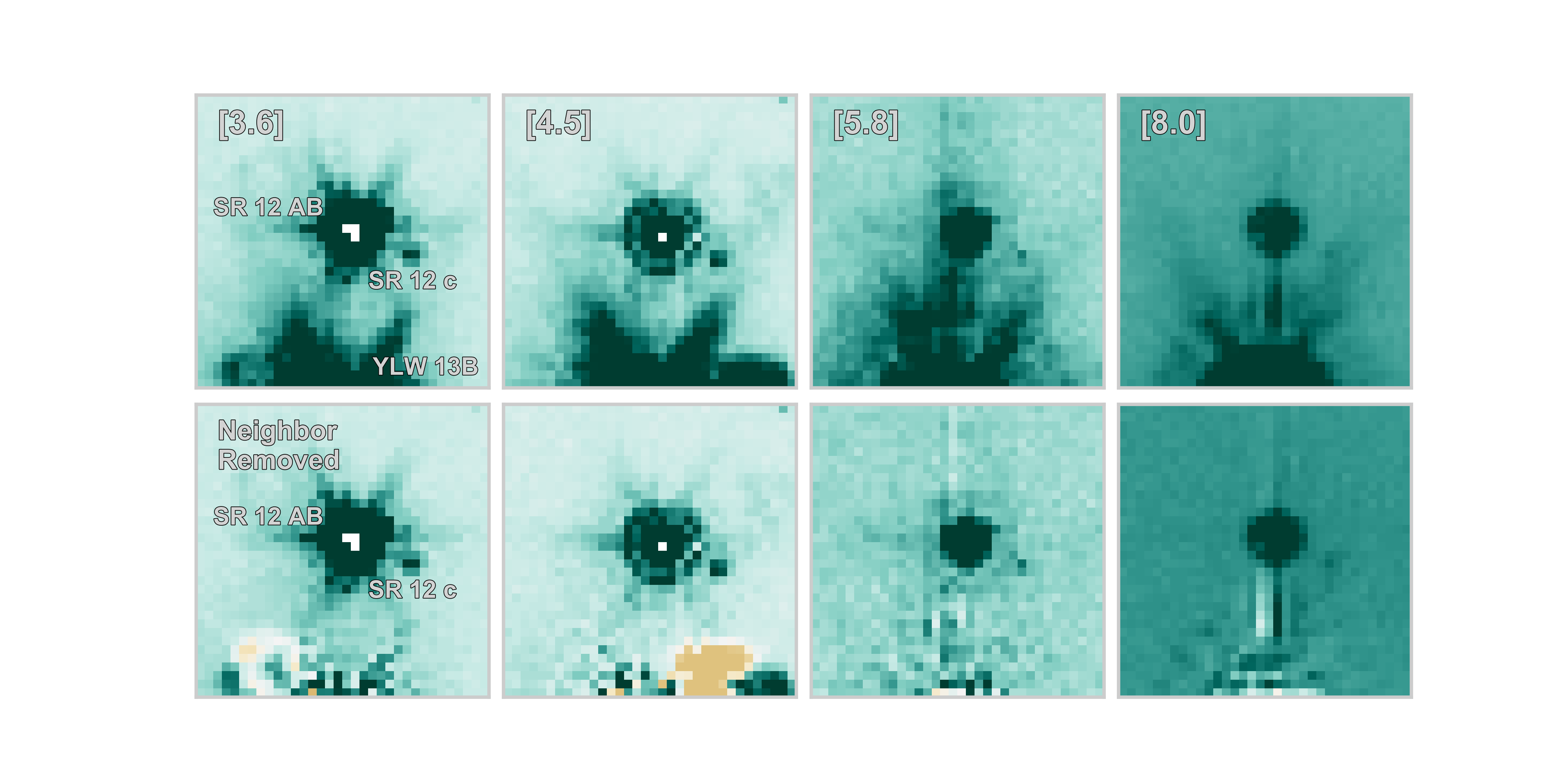}
    \caption{Individual IRAC images of SR 12 before (top row) and after (bottom row) YLW 13B, a nearby young stellar object, is removed. High-dynamic range PSFs were used to model the bright PSF wings of YLW 13B which were then subtracted off to minimize contamination when determining the system parameters of SR 12.}
    \label{fig:SR12n}
\end{figure*}

Some members of the sample have nearby neighbors with flux that could influence the results of the pipeline fit. The neighbors of DH Tau, 2MASS J$04414565$+$2301580$ and 2MASS J$04414489$+$2301513$ (hereafter 2M0441 A and 2M0441 B), and CHXR 73 are within $15\arcsec$ of the primary centroid and unsaturated. We use the same PSF model described above to fit and subtract each neighbor within each individual IRAC image prior to being put through the pipeline. SR 12 is $\sim$25$\arcsec$ away from a bright and saturated young stellar object, 2MASS J16272146--2441430 (YLW 13B). Although this object is well outside of the pipeline fitting region,the wings of its flux can still affect the PSF-fitting results. For this system we use the high-dynamic range PSF from \cite{marengo06} to model this bright neighbor and subtract off its contaminating flux (Figure \ref{fig:SR12n}).

After the MCMC runs, stacked residual images are created by combining individual residual images after the primary PSF has been subtracted, placing each on a final grid with a pixel scale five times smaller than the original IRAC pixel scale of $1\farcs22$, shifting to a common origin, and rotating so that north is up and east is left. PSF subtraction occurs on the original data, not on mosaiced or subsampled images, because of the complicated nature of the IRAC PSF and because subsampling the images prior to PSF-fitting would introduce covariance between adjacent pixels. We perform aperture photometry on these subsampled stacked residuals images to determine detection limits around each primary. We use apertures with radii equal to the FWHM in each channel ($1\farcs66$, $1\farcs72$, $1\farcs88$, $1\farcs98$). The FWHMs are larger than the IRAC pixel scale ($1\farcs22$), thus all covariant pixels contribute to the measured aperture flux.

To evaluate the sensitivity of our PSF-fitting framework to substellar companions in the IRAC images of our sample, we performed aperture photometry on the stacked images before and after PSF-subtraction. We measured the flux inside 100 randomly drawn apertures of radius 1 FWHM at $\mathrm{FWHM}/4$ ($0\farcs42$, $0\farcs43$, $0\farcs47$, $0\farcs50$) intervals radially outward from the primary star. The mean and standard deviation of these fluxes is used to determine the limiting flux and is then converted into \textit{Spitzer}/IRAC magnitudes to obtain 4-$\sigma$ limits. With $\sim$36, 34, 28, and 25 independent apertures in a search radius of $10\arcsec$ around a primary star, the probability of measuring a spurious $>$4-$\sigma$ signal is 0.003\%.

We then convert our detection limit curves into mass detection limits using the BT-Settl evolutionary models of \cite{allard12} at the reported literature ages and \textit{Gaia} parallactic distances of our sample systems.

For a given target, the companion height above (or below) the 4-$\sigma$ detection limit at that radius can be used to infer the systematic uncertainty due to residual primary-PSF structure in our modeling framework photometry. For example, if photometry measured for a companion is equal to the 4-$\sigma$ limit, its systematic flux uncertainty would be 25\%, or $\sim$0.24 mag, while a 5-$\sigma$ detection would have a systematic flux uncertainty of 20\%, or $\sim$0.20 mag. For all contrast and photometry measurements hereafter, we list this systematic uncertainty in addition to the statistical uncertainty from our MCMC fits.

\section{Results} \label{sec:results}
\subsection{Detections}
Our reprocessing of the IRAC images yielded detections in one or more filters for eight out of our sample of nine substellar companions. The one system whose companion was not detected, GJ 504, had the brightest primary ($K_s$=4.03 mag) and largest expected contrast ($>$12 mag), but we are still able to assess a robust upper limit. We present the final system parameters as determined by our pipeline in Table \ref{doublefit_tab}. The contrasts reported are marginalized values of the parameters as measured by our MCMC fits and we list both statistical and systematic uncertainties. Further analysis we perform with this output adds the uncertainties in quadrature. The projected separations and position angles reflect the input priors from previous adaptive optics imaging such that the information in the \textit{Spitzer}/IRAC images is entirely devoted to measuring companion contrast. IRAC magnitudes for the primary stars and substellar companions are calculated from the PSF models, assuming the median MCMC fit parameters, and are included in Table \ref{tab:phot}.

\begin{deluxetable*}{lccccccc}
\tabletypesize{\footnotesize}
\tablecaption{Best-Fit System Properties of Detected Companions}
\tablehead{\colhead{2MASS} & \colhead{Other Name} & \colhead{Separation} & \colhead{Position Angle} & \colhead{$\Delta [3.6]$} & \colhead{$\Delta [4.5]$} & \colhead{$\Delta [5.8]$} & \colhead{$\Delta [8.0]$}\\
\colhead{(Primary)}& \colhead{(Companion)} & \colhead{(arcsec)} & \colhead{(deg)} & \colhead{(mag)} & \colhead{(mag)} & \colhead{(mag)} & \colhead{(mag)}}
\startdata
J04294155+2632582   & DH Tau B          & $2.22\pm0.18$ & $137.30\pm1.5$& $5.74\pm0.24\pm0.56$	& $5.30\pm0.17\pm0.42$	& $4.79\pm0.16\pm0.31$	& $4.38\pm0.15\pm0.19$ \\
J04414565+2301580   & 2M0441 Bab        & $12.35\pm0.01$& $237.4\pm0.1$ & $2.67\pm0.01\pm0.01$ & $2.43\pm0.01\pm0.01$ & $2.16\pm0.01\pm0.01$ & $1.78\pm0.01\pm0.01$ \\
J06191291--5803156 	& AB Pic b			& $5.52\pm0.09$ & $175.4\pm0.3$ & $6.34\pm0.06\pm0.08$ & $5.98\pm0.07\pm0.09$ & $5.65\pm0.26\pm0.09$ & $5.58\pm0.16\pm0.20$ \\
J11062877--7737331  & CHXR 73 b			& $1.24\pm0.03$ & $228.5\pm3.8$	& $3.63\pm0.04\pm0.12$ & $3.79\pm0.06\pm0.04$ & ...           & $2.92\pm0.12\pm0.08$ \\
J16093030--2104589  & 1RXS J1609 b      & $2.14\pm0.11$ & $27.0\pm0.6$  & ...           & $6.04\pm0.15\pm0.17$ & ...           & ...           \\
J16271951--2441403 	& SR 12 c			& $8.62\pm0.05$ & $164.8\pm0.6$ & $5.83\pm0.07\pm0.08$ & $5.08\pm0.03\pm0.06$ & $5.05\pm0.07\pm0.13$ & $4.16\pm0.03\pm0.12$ \\
J16311501--2432436 	& ROXs 42B			& $1.17\pm0.03$ & $263.6\pm4.8$	& ...           & $5.08\pm0.56\pm0.24$	& ...           & $4.55\pm0.11\pm0.10$ \\
J21185820+2613500   & HD 203030 b		& $12.025\pm0.004$ & $108.69\pm0.02$    & $8.27\pm0.01\pm0.63$ & $7.88\pm0.01\pm0.27$	& $6.97\pm0.02\pm0.20$ & $6.84\pm0.02\pm0.21$
\enddata
\tablecomments{If an entry in $\Delta m$ is missing, the companion was not detected in that channel.}
\label{doublefit_tab}
\end{deluxetable*}

\begin{figure*}
    \centering
    \includegraphics[trim=1.5cm 2cm 1.15cm 2cm,clip,width=\textwidth]{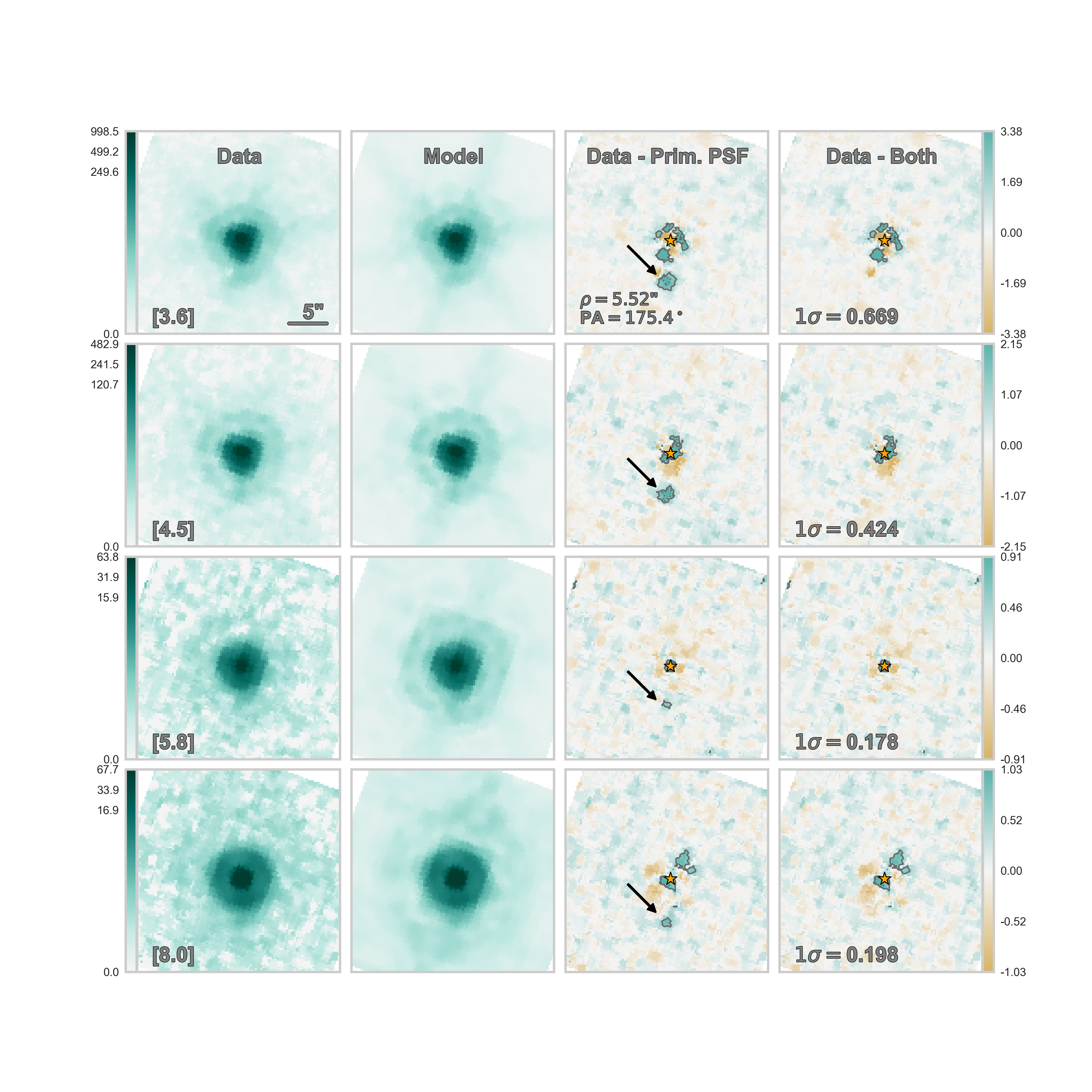}
    \caption{Stacked images of AB Pic across all four IRAC channels (rows) after it has gone through the PSF-fitting pipeline. All fits were conducted within the CBCD images at the native plate scale, but to convey the full data set, the images here were generated by combining individual frames after they had been re-scaled to $0\farcs24$/pixel ($\sim$5$\times$ smaller than the original IRAC pixel scale), shifted to a common origin, and rotated so that north is up and east is left. Columns 1 and 2 show the original IRAC data of AB Pic and the median two-source PSF model, respectively, displayed with a logarithmic color scale (leftmost color bar). Column 3 shows the residuals left behind after only the primary PSF model is subtracted from the data. Column 4 shows the residuals left behind after the two-source PSF model is subtracted from the data.  Both Columns 3 and 4 are displayed with a linear color scale (rightmost color bar) and 3- and 5-$\sigma$ contours overlaid with solid and dotted lines, respectively. The standard deviation of the pixel values is displayed in the lower left-hand corner of Column 4 in units of DN/s. After subtracting the primary star PSF, a statistically significant positive residual is seen at the expected position of AB Pic b. This residual disappears after subtracting the best-fit system PSFs, indicating that it is a robust detection across all IRAC filters.}
    \label{fig:abpic}
\end{figure*}

In Figure \ref{fig:abpic}, we present example pipeline results for an individual system, AB Pic. We show stacked images of the original data and final system model as well as stacked residuals images after the PSF models are subtracted. After subtracting the primary star PSF, a statistically significant positive residual is seen at the expected position of AB Pic b. This residual disappears after subtracting the best-fit system PSFs, indicating that it is a robust detection across all IRAC filters.

\begin{figure*}
    \centering
    \includegraphics[trim=1.0cm 2cm 0.5cm 1.75cm,clip,width=\textwidth]{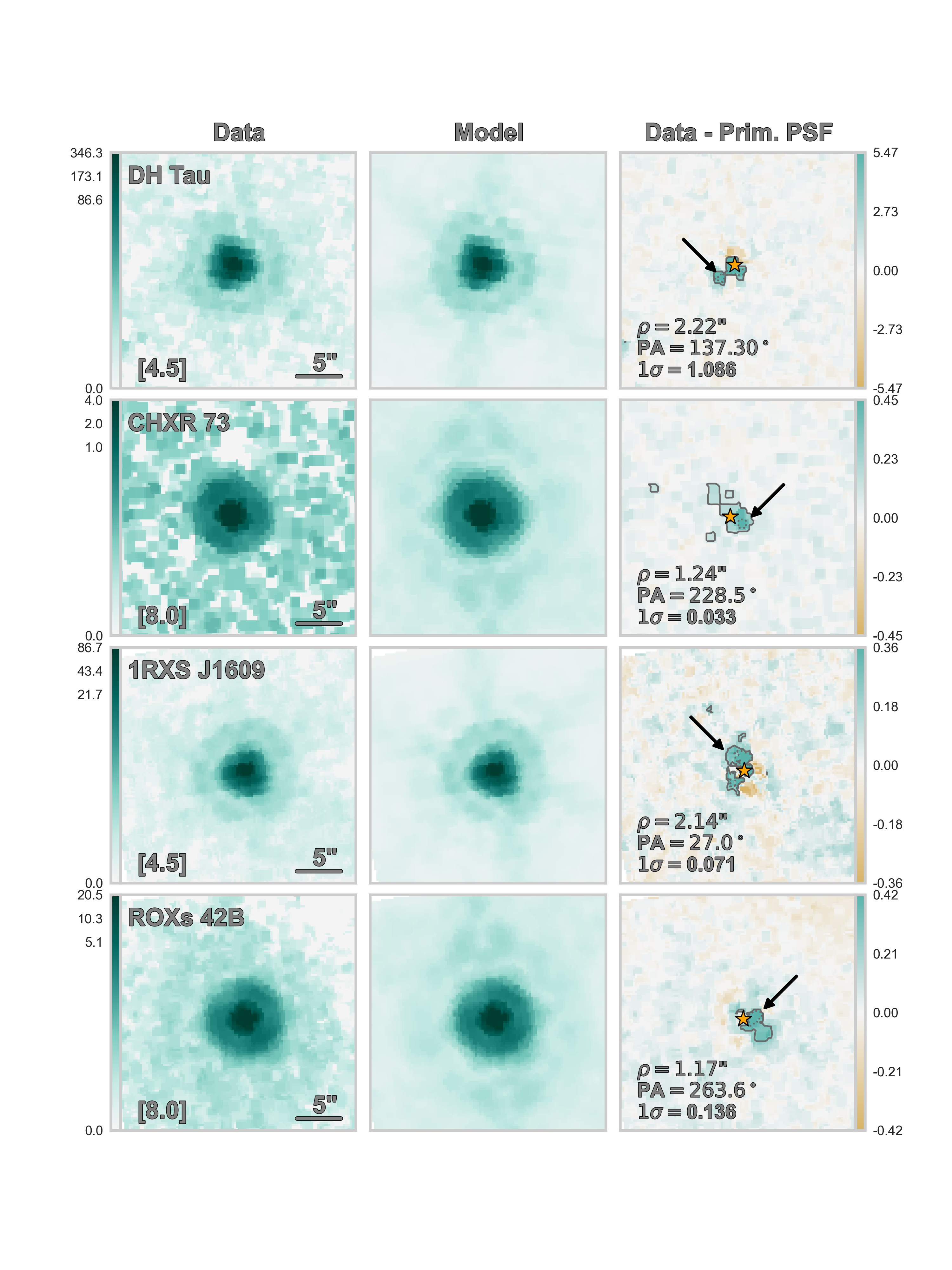}
    \caption{Stacked images of DH Tau, CHXR 73, 1RXS J1609, and ROXs 42B, the other four systems besides AB Pic (shown in Figure \ref{fig:abpic}) with companions that are newly resolved in this work. Columns 1 and 2 show the original IRAC data and the median two-source PSF model, respectively, displayed with a logarithmic color scale (leftmost color bar). Column 3 shows the residuals left behind after only the primary PSF model is subtracted from the data. Column 3 is displayed with a linear color scale (rightmost color bar) with 3- and 5-$\sigma$ contours overlaid with solid and dotted lines, respectively. For each panel north is up and east is left.}
    \label{fig:comp_examples}
\end{figure*}

Not all companions were detected in every IRAC channel. Generally, companions were not detected or had less constrained photometry in Channel 3 (5.8 $\mu$m), suggesting a possible PSF mismatch between templates and data in that channel. CHXR 73 b was detected in Channels 1, 2, and 4. ROXs 42B b was detected in Channels 2 and 4. 1RXS J160929.1--210524 b (hereafter 1RXS J1609 b) was detected only in Channel 2. These three objects had the smallest projected separations ($1\farcs2-2\farcs2$) of the sample. ROXs 42B b and 1RXS J1609 b also had the largest $K_s$-band contrasts which could explain the difficulty of detection in the other IRAC channels. Our measured photometry in Channel 4 (8.0 $\mu$m) of ROXs 42B b suggests it may have a long-wavelength excess, making its detection easier. In Figure \ref{fig:comp_examples}, we show stacked residuals images after the primary PSF has been subtracted, highlighting the companion detection in either Channel 2 or Channel 4 for these systems, as well as DH Tau.

\begin{figure}
    \centering
     \includegraphics[trim=1.5cm 0.5cm 0.25cm 0cm,clip,width=0.45\textwidth]{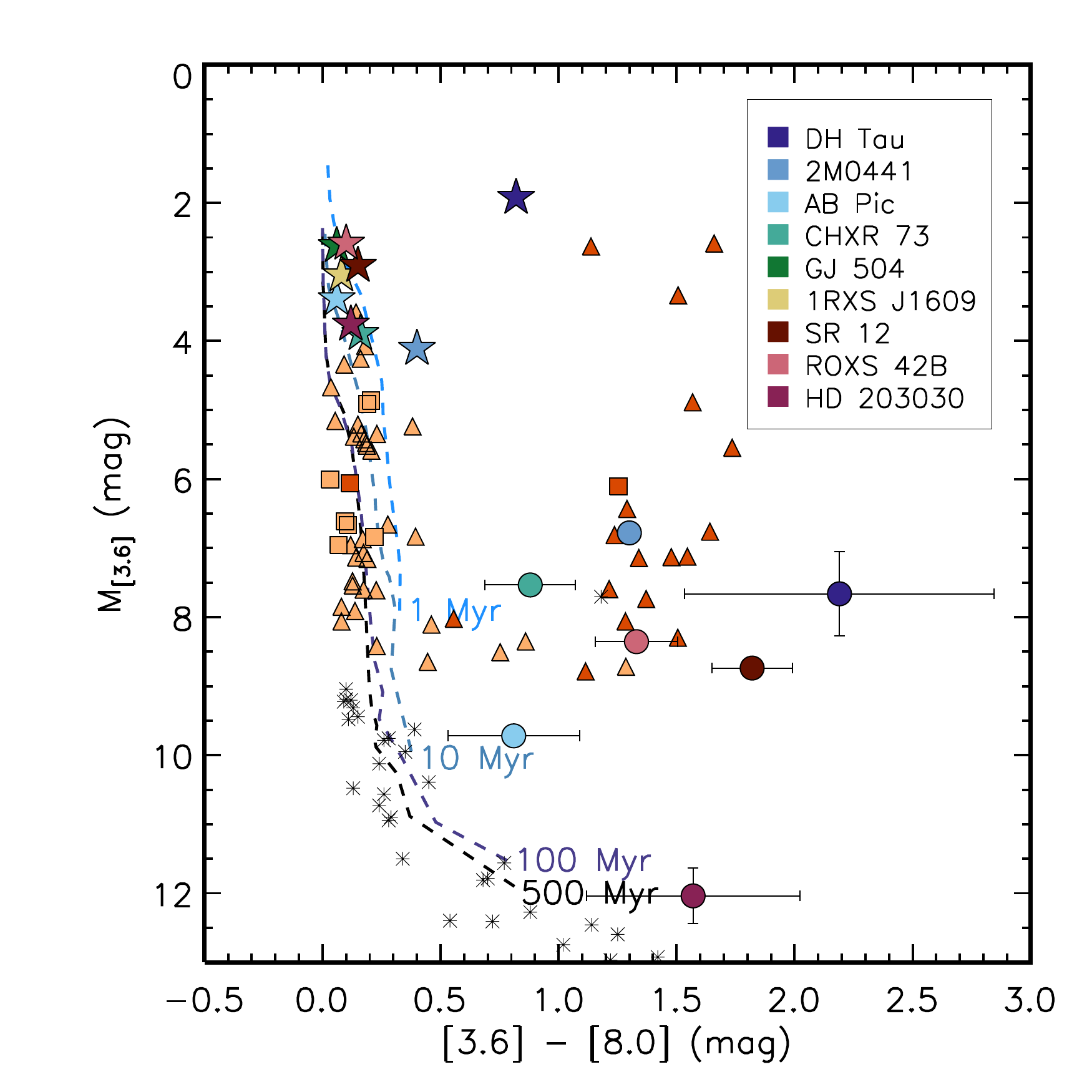}
    \caption{Color-magnitude diagram for our sample detected in both Channels 1 (3.6 $\mu$m) and 4 (8.0 $\mu$m). For comparison, we include young $>$M7 brown dwarfs members of the Taurus (triangles; \citealt{esplin17}) and Upper Scorpius (squares; \citealt{luhman20}) star-forming regions. Orange symbols represent disk-free members while red symbols denote disk-bearing members. We also include field brown dwarfs from \citet{dupuy12}, indicated as asterisks. $M_{[3.6]}$ was determined from \textit{Gaia} EDR3 parallactic measurements \citep{bailer-jones21} of each system primary. The primary components are indicated as filled stars while substellar companions are indicated as filled circles. Also displayed are the intrinsic photospheric [3.6]-[8.0] colors from BT-Settl models of \cite{allard12} at 1, 10, 100, and 500 Myr (dashed lines). Not shown are the companions to GJ 504 and 1RXS J1609 which were not detected in either channels 1 or 4. ROXs 42B b was not detected in Channel 1 by our pipeline but shown here as an $L^{\prime}$-band detection from \cite{kraus14}. The companions of our sample appear to be significantly redder than the BT-Settl isochrones, in line with previous comparisons between young and old free-floating brown dwarfs \citep{dupuy12,liu16}. The mid-infrared colors of the companions are similar to disk-bearing free-floating brown dwarfs in young star-forming regions.}
    \label{fig:cmd}
\end{figure}

In Figure \ref{fig:cmd}, we present a color-magnitude diagram of $M_{[3.6]}$ vs.~[3.6]--[8.0] color for the nine primaries and seven companions that were detected in those filters. We also show the intrinsic photospheric mid-infrared color-magnitude sequences from the BT-Settl models of \cite{allard12} for 1, 10, 100, and 500 Myr. Typically an object with [3.6]--[8.0] color significantly redder than the intrinsic photospheric isochrone on this diagram is interpreted as excess emission due to a disk. Based on this criterion, two primaries and seven companions appear red and may harbor circum(sub)stellar disks, but we will explore whether a more nuanced disk criterion is needed in Section \ref{discussion}.

We detect the photospheres of AB Pic b, CHXR 73 b, and HD 203030 b while companions with significant [3.6]--[8.0] color excess are DH Tau B, SR 12 c, and ROXs 42B b. Although 1RXS J1609 b was not detected in Channels 1, 3, or 4, an SED fit of literature photometry and our Channel 2 measurement indicate we detected its photosphere (see Section \ref{sec:sedfits}).

2M0441 AB actually comprise a quadruple system consisting of two bound low-mass binaries (\citealt{todorov10,todorov14,bowler15}, and references therein). Mid-infrared excess has been identified for both pairs \citep{luhman10,adame11,bulger14}, indicating at least one component of each binary harbors a circum(sub)stellar disk. We readily confirm this excess with our pipeline in the \textit{Spitzer} images but determining the mid-infrared flux contributions from the individual components of 2M0441 B is beyond the scope of this paper.

\begin{figure*}
    \centering
    \includegraphics[angle=90,trim=1.25cm 2.0cm 1.15cm 1.75cm,clip,width=\textwidth]{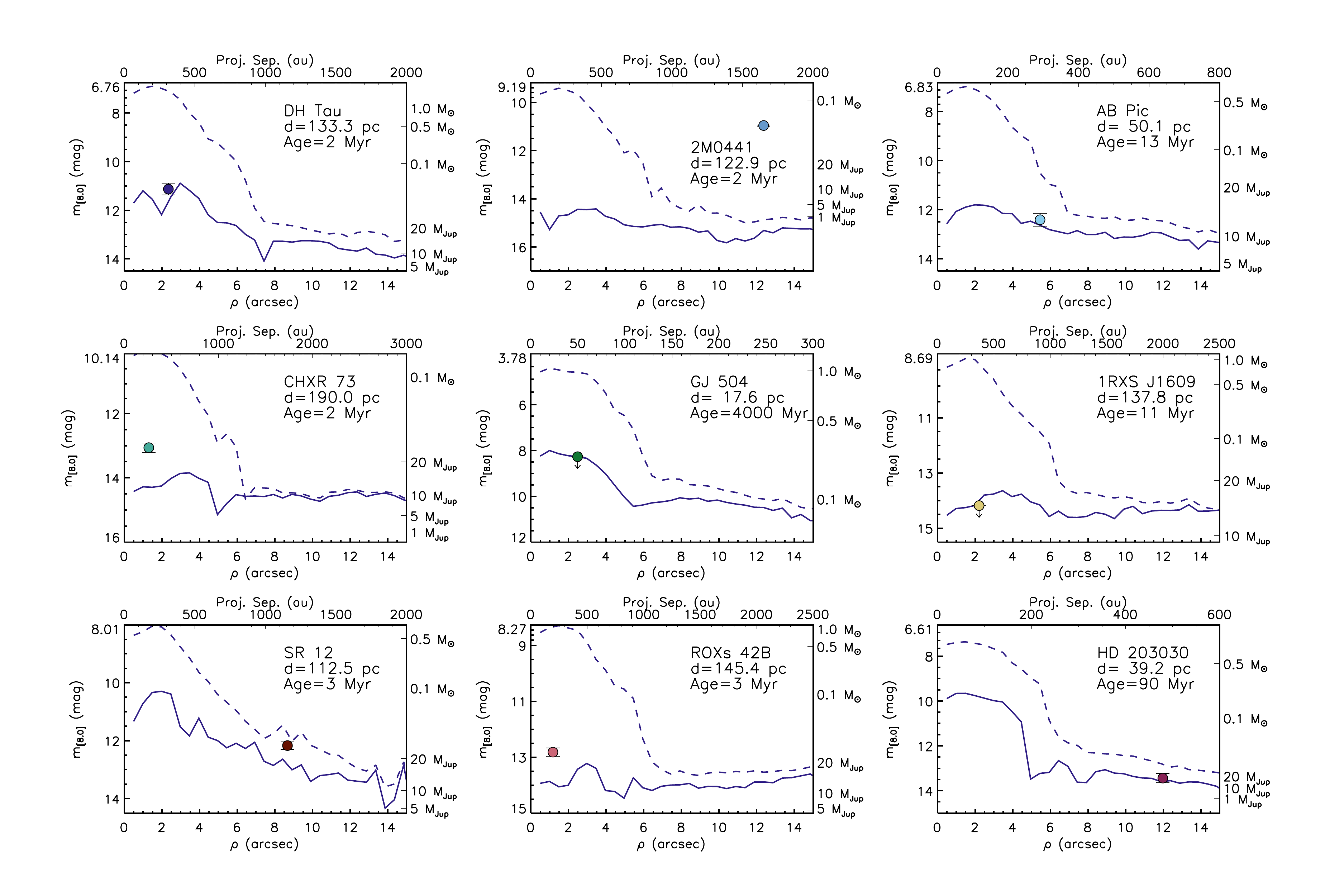}
    \caption{Contrast limits determined from the stacked IRAC Channel 4 images of our sample. The dashed blue line indicates the contrast curves prior to PSF subtraction as a function of separation from the primary in arcseconds. The solid blue line indicates the corresponding contrast curve after the median two-source PSF model has been subtracted. The top of each leftside y-axis lists the primary magnitude as measured by the pipeline. The contrast curves are presented in terms of apparent magnitudes as well as mass calculated by using the \textit{Gaia} EDR3 parallax distance estimates of \cite{bailer-jones21}, literature age determinations, and BT-Settl isochrones \citep{allard12}.}
    \label{fig:dl}
\end{figure*}

\begin{figure*}
    \centering
    \includegraphics[width=\textwidth]{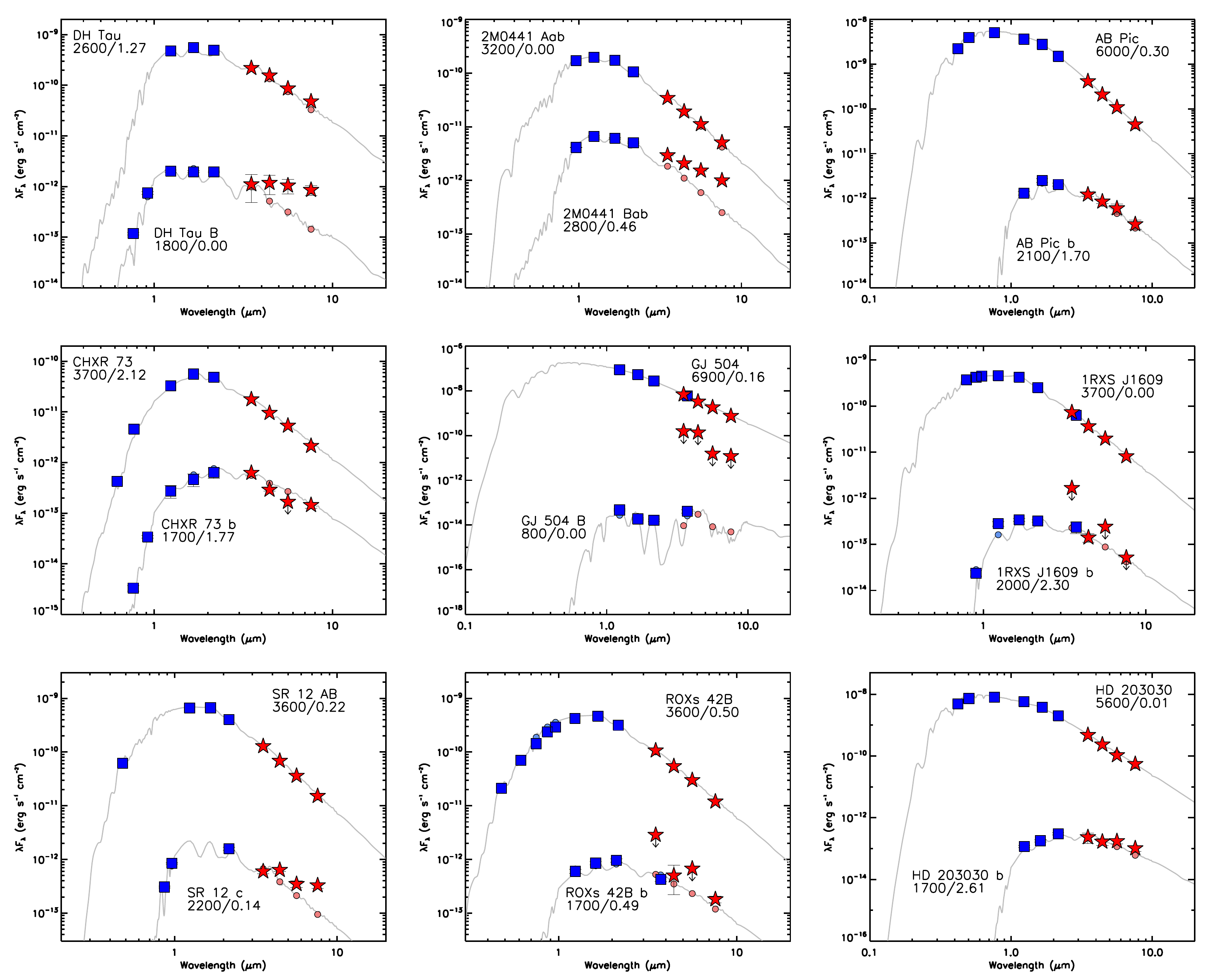}
    \caption{Spectral energy distributions of our sample. We fit system components with solar metallicity BT-Settl model atmospheres \citep{allard12} and fit $E(B-V)$ individually as a free parameter. SED-fitting results ($T_{\rm eff}$/$E(B-V)$) for each component of our wide companion systems are plotted near the reddened best-fit model. Literature photometry for the systems is plotted as blue squares while \textit{Spitzer}/IRAC photometry from this work is indicated as red stars. Synthetic photometry of the best fit model is also plotted as blue circles for the literature filters used or red circles at the IRAC channels. Specific filters and photometry used in our SED fits for the entire sample are listed in Table \ref{tab:phot}. SED-fitting results for the entire sample, including the $\chi^2_{\nu}$ of each fit and $[8.0]_{mod}-[8.0]_{obs}$ magnitude excess, are listed in Table \ref{tab:sedfit}.}
    \label{fig:sed}
\end{figure*}

\subsection{Detection Limits and Mass Sensitivity}
Our PSF-fitting results yield sensitive upper limits on the companions that we did not detect, as well as for the presence of additional companions in these systems. We present the contrast and mass limits reached in the PSF-subtracted images as a function of radial separation in Tables \ref{tab:contrast_limits} and \ref{tab:mass_limits}, and show the Channel 4 detection limits in Figure \ref{fig:dl} (See Section \ref{data_analysis} for the details of the detection limit calculation).

Our detection limit curves show that prior to PSF subtraction detectable companion contrasts plateau past $8\arcsec$, corresponding to projected physical separations of $\sim$150 au for the closest sample member (17.5 pc; GJ 504) and $\sim$1600 au for the furthest (191.0 pc; CHXR 73). Within $8\arcsec$, detectable contrasts improve by as much as 7 mag in Channel 1 and 5.5 mag in Channel 4.

While we do not detect the companion of GJ 504 in the images, we can place a limit on its $L-$[8.0] color using previous AO imaging results in the $L$-band from \cite{skemer16} and our work. We find $L-[8.0]<8.43$ mag.

\subsection{SED Fits} \label{sec:sedfits}

Optical and near-infrared photometry from the literature can be used with our new \textit{Spitzer}/IRAC mid-infrared photometry to analyze the SEDs of our sample systems. We fit system components with solar metallicity BT-Settl model atmospheres \citep{allard12} spanning effective temperatures between 1000 and 7000 K ($\Delta T_{\mathrm{eff}}=100$ K) fixed at either $\log g=3.5$ (2M0441 B, CHXR 73, and ROXS 42B b) or $\log g=4.0$ which is appropriate for young dwarfs according to BT-Settl evolutionary models. We convolve the model atmospheric spectra with filter transmission profiles to generate synthetic photometric measurements and find the $\chi^2$-minimizing scale factor between the model photometry and the observed photometry for each object. We also fit $E(B-V)$ as a free parameter using the extinction curve of \cite{fitzpatrick99} in steps of 0.01 mag.

In Figure \ref{fig:sed}, we show the SED fits for the systems in our sample. We show our new \textit{Spitzer}/IRAC photometry as red stars and the literature photometry used in our fits as blue squares. We also include the best-fit BT-Settl model for the primary and companion in gray. In Table \ref{tab:phot}, we list all the photometry used in each SED fit for each system, in addition to our measured photometry from the pipeline parameters. In Table \ref{tab:sedfit}, we summarize the properties found from our fits. We discuss the SED fits of specific systems in Section \ref{system_notes}, and interpret potential companion mid-infrared excesses further in Section \ref{discussion}.

\subsection{Notes on Two Individual Systems}
\label{system_notes}
Our reprocessing of the IRAC images yielded detection of all nine primaries and eight substellar companions, five of which have not been resolved in the IRAC filters. We describe two systems in more detail in the following sections.

\subsubsection{DH Tau} 
DH Tau is a protoplanetary disk host with spectral type M1 \citep{herbig77,watson09}. It is an actively accreting classical T Tauri star with previously detected mid-infrared excess \citep{valenti93,meyer97,luhman06c,luhman10}. DH Tau hosts a substellar companion at projected separation $\rho=2.3$ ($\sim$310 au at its \textit{Gaia} distance of $\sim$135 pc; \citealt{itoh05,bailer-jones21}). The companion mass was initially estimated to be $\sim$30--50 $M_{\mathrm{Jup}}$ but comparison of its bolometric luminosity to newer evolutionary models revealed a lower mass of $\sim$11 $M_{\mathrm{Jup}}$ \citep{luhman06c,kraus14,bowler16}, closer to the planet--brown dwarf boundary. Hydrogen emission lines and a UV continuum excess indicate active accretion onto DH Tau B \citep{zhou14,bonnefoy14} but emission from the circum(sub)stellar disk has not been detected \citep{wu20}.

We resolve DH Tau B and measure its mid-infrared photometry in all IRAC channels (see Fig \ref{fig:comp_examples} for Channel 2 detection). As we described in \ref{sec:sedfits} and show in Figure \ref{fig:sed}, we use \textit{Hubble Space Telescope} (\textit{HST}) optical \citep{zhou14}, Two Micron All Sky Survey (2MASS) near-infrared \citep{cutri03}, and the \textit{Spitzer}/IRAC 3.6 $\mu$m photometry measured in this work to analyze the SEDs of DH Tau A and B. The best-fitting model for DH Tau A is $T_{\mathrm{eff}}=2600\pm100$ K and $E(B-V)=1.27\pm0.06$ mag and for DH Tau B, the best-fitting model is $T_{\mathrm{eff}}=1800\pm50$ K and $E(B-V)=0.00\pm0.03$ mag. The 8 $\mu$m photometry for DH Tau B disagrees with the best-fitting model at 3.8-$\sigma$. We find the mid-infrared color of DH Tau B to be [3.6]--[8.0]$=2.19\pm0.66$ mag which is discrepant with the \cite{luhman10} empirical color of an M9 dwarf atmosphere at the 2.7-$\sigma$ level. Using the dereddened $K$-band photometry from \citet{itoh05} and our Channel 4 detection, DH Tau B's infrared color is $K$--[8.0]$=2.93\pm0.24$ mag which is discrepant from an M9 atmosphere at at the 8.0-$\sigma$ level. This red color indicates a clear mid-infrared excess consistent with presence of circum(sub)stellar disk.

We use our 3.6 $\mu$m photometric measurement, DH Tau's \textit{Gaia} parallactic distance \citep[133.3 pc;][]{bailer-jones21}, and Taurus's adopted age of $\tau\sim2$ Myr to estimate the mass of DH Tau B to be $M=17\pm6$ $M_\mathrm{Jup}$, consistent with previous mass determinations \citep[$M=18\pm4$ $M_{\rm Jup}$;][]{kraus14}.

\subsubsection{AB Pic}

AB Pic is a K2 star originally considered a member of the Tucana-Horologium association ($\tau\sim40$ Myr; \citealt{song03,kraus14b,bell15}). \cite{torres08} later re-assessed AB Pic to be a member of Carina, another young moving group (YMG) with age $\tau\sim30$ Myr \citep[e.g.,][]{bell15,miret-roig18}. Recently, \cite{booth2021} has revised the age of Carina to be younger and only 13 Myr.

\cite{chauvin05} observed AB Pic to host a planetary-mass companion, AB Pic b, at projected separation $5\farcs5$ or 275 au at its 50 pc distance. Near-infrared spectroscopic observations measure the spectral type of the wide companion to be L0-L1 \citep{bonnefoy14}. Near-infrared spectroscopy of the companion has not detected emission line accretion indicators \citep{bonnefoy14}, nor has the companion been detected at wavelengths longer than $L^\prime$ filter \citep{rameau13,perez19}.

We detect AB Pic b in all four IRAC channels, finding its mid-infrared color to be $[3.6]-[8.0]=0.81\pm0.27$ mag. AB Pic b is consistent with younger L0 photospheres to within 1-$\sigma$ \citep{luhman10}. Using near-infrared photometric measurements from \cite{chauvin05} and the mid-infrared photometry found in this work, we analyze the SED of AB Pic b similarly as we described previously for DH Tau B (see Figure \ref{fig:sed}) allowing $E(B-V)$ to range up to 2.0 mag. The best-fitting models for AB Pic A and b have $T_{\mathrm{eff}}=6000\pm100$ K and $T_{\mathrm{eff}}=2100\pm100$ but with discrepant $E(B-V)$ of $0.30\pm0.02$ mag and $1.70\pm0.19$ mag, respectively. The amount of reddening for the primary is consistent with previous measurements \citep[$0.27\pm0.02$ mag;][]{vanBelle09}. The significantly higher value found for AB Pic b could indicate the presence of an as of yet unresolved circum(sub)stellar disk, though the observations allow a substantial range of possible values. Our observed 8 $\mu$m flux disagrees with the model photosphere at the 0.7-$\sigma$ level. Fixing the companion color excess to agree with the primary at $E(B-V)=0.30$ mag results in a cooler best-fitting model with $T_{\mathrm{eff}}=1600$ K. For this model photosphere, the observed 8 $\mu$m flux from our work disagrees only at the 0.3-$\sigma$ level. We therefore conclude that there is not yet compelling evidence that AB Pic b hosts a disk or shows a mid-infrared excess.

Using our 3.6 $\mu$m photometric measurement, the \textit{Gaia} parallactic distance \citep[50.1 pc;][]{bailer-jones21}, and revised age of Carina ($\tau\sim13$ Myr), we estimate the mass of AB Pic b to be $M=11\pm1$ $M_{\rm Jup}$. This new mass estimate is $\sim$2 $M_{\rm Jup}$ lower than if an age of 30 Myr were assumed for AB Pic and places the companion firmly below the deuterium-burning limit.

\subsection{Companions with Previous IRAC Photometric Measurements}

\subsubsection{2M0441 AB}
\cite{esplin14} report IRAC photometry of $m_{[4.5]}=9.48\pm0.02$ mag, $m_{[5.8]}=9.37\pm0.03$ mag, and $m_{[8.0]}=9.22\pm0.03$ mag for 2M0441 A in Channels 2--4 (Channel 1 is saturated). For 2M0441 B, they report $m_{[3.6]}=12.26\pm0.02$ mag, $m_{[4.5]}=11.88\pm0.02$ mag, $m_{[5.8]}=11.53\pm0.03$ mag, and $m_{[8.0]}=11.00\pm0.03$ mag. Our pipeline IRAC photometry agrees with these prior measurements within the error bars, and through our PSF-fitting procedure that masks saturated pixels, we are able to recover a Channel 1 magnitude for the the primary that is consistent with its overall SED shape.

\subsubsection{HD 203030}
\cite{miles-paez17} observed HD 203030 with \textit{Spitzer}/IRAC (Program ID 40489) at two distinct epochs to utilize roll subtraction to obtain photometry of HD 203030 b that was minimally contaminated by its host bright stellar halo. No photometry is reported for HD 203030, but they do report IRAC photometry of $m_{[3.6]}=14.99\pm0.02$ mag, $m_{[4.5]}=14.73\pm0.02$ mag, $m_{[5.8]}=14.39\pm0.05$ mag, and $m_{[8.0]}=14.15\pm0.04$ mag for HD 203030 b. We initially fit the first epoch long-exposure images of HD 203030 and found our Channel 1 pipeline photometry of $m_{[3.6]}=15.01\pm0.40$ mag agreed with their measurement, but we measured higher fluxes for all other channels ($m_{[4.5]}=14.63\pm0.26$ mag, $m_{[5.8]}=13.84\pm0.20$ mag, and $m_{[8.0]}=13.44\pm0.21$ mag). No ground-based photometry at wavelengths greater than 3 $\mu$m have been reported in the literature.

To explore this discrepancy further, we use our pipeline infrastructure to fit all of the IRAC images available simultaneously for each exposure time, as well as the individual epochs separately. In the 1 s exposures, the primary is not saturated and we measure its photometry to be $m_{[3.6]}=6.64\pm0.02$ mag, $m_{[4.5]}=6.68\pm0.02$ mag, $m_{[5.8]}=6.63\pm0.02$ mag, and $m_{[8.0]}=6.63\pm0.02$ mag across the IRAC channels for the first epoch, and $m_{[3.6]}=6.66\pm0.02$ mag, $m_{[4.5]}=6.68\pm0.02$ mag, $m_{[5.8]}=6.64\pm0.02$ mag, and $m_{[8.0]}=6.63\pm0.02$ mag for the second. The IRAC photometry is consistent with the Band 1 (3.5 $\mu$m), 2 (4.6 $\mu$m), and 3 (12.0 $\mu$m) mid-infrared photometry reported in the AllWISE catalog \cite{cutri14} ($W1=6.66\pm0.07$, $W2=6.63\pm0.02$, $W3=6.63\pm0.02$). In the 26.8 s exposures, our pipeline measures $m_{[3.6]}=6.73\pm0.02$ mag, $m_{[4.5]}=6.76\pm0.02$ mag, $m_{[5.8]}=6.88\pm0.02$ mag, and $m_{[8.0]}=6.61\pm0.02$ mag across the IRAC channels for the first epoch, and $m_{[3.6]}=6.80\pm0.02$ mag, $m_{[4.5]}=6.79\pm0.02$ mag, $m_{[5.8]}=6.69\pm0.02$ mag, and $m_{[8.0]}=6.61\pm0.02$ mag for the second, suggesting a systematic uncertainty of 0.1--0.2 mag for the brightness of the primary when its PSF core is saturated. This variation is still smaller than the $\sim$0.5--0.7 mag discrepancy between our photometric measurements for the companion and those of \cite{miles-paez17}, indicating that if there is an issue in our analysis, it comes when fitting the companion and after the primary PSF has been subtracted.

The intrinsic photospheric colors of late-M and L dwarfs are typically determined by combining photometric observations with parallax measurements \citep[e.g.,][]{patten06,luhman10,filippazzo15,faherty16}. A L7.5 field dwarf should have $K_s-[3.6]=1.13$ mag, $[3.6]-[4.5]=-0.02$ mag, $[4.5]-[5.8]=0.32$ mag, and $[5.8]-[8.0]=0.23$ mag in the IRAC channels, according to \cite{dupuy12}. Both $[3.6]-[4.5]$ colors measured for HD 203030 b by \cite{miles-paez17} and in this work are $\sim$0.3--0.4 mag redder than expected, although consistent with younger planetary-mass objects \citep[e.g.,][]{filippazzo15,faherty16,liu16}. In the other channels, the colors measured by \cite{miles-paez17} agree with a field L7.5 photosphere, while our photometry continues to be significantly redder. \cite{miles-paez17} measure $[3.6]-[8.0]=0.84\pm0.04$ which is $\sim$0.3 mag redder than expected but still within the upper envelope of the rms scatter of the \cite{dupuy12} sample. We measure  $[3.6]-[8.0]=1.57\pm0.45$ mag. This color excess could potentially indicate the presence of a circum(sub)stellar disk if confirmed, but given the disagreement with \cite{miles-paez17}, such an interpretation should be treated with caution.

\begin{figure*}
    \centering
    \includegraphics[trim=1.5cm 2cm 1.15cm 2cm,clip,width=\textwidth]{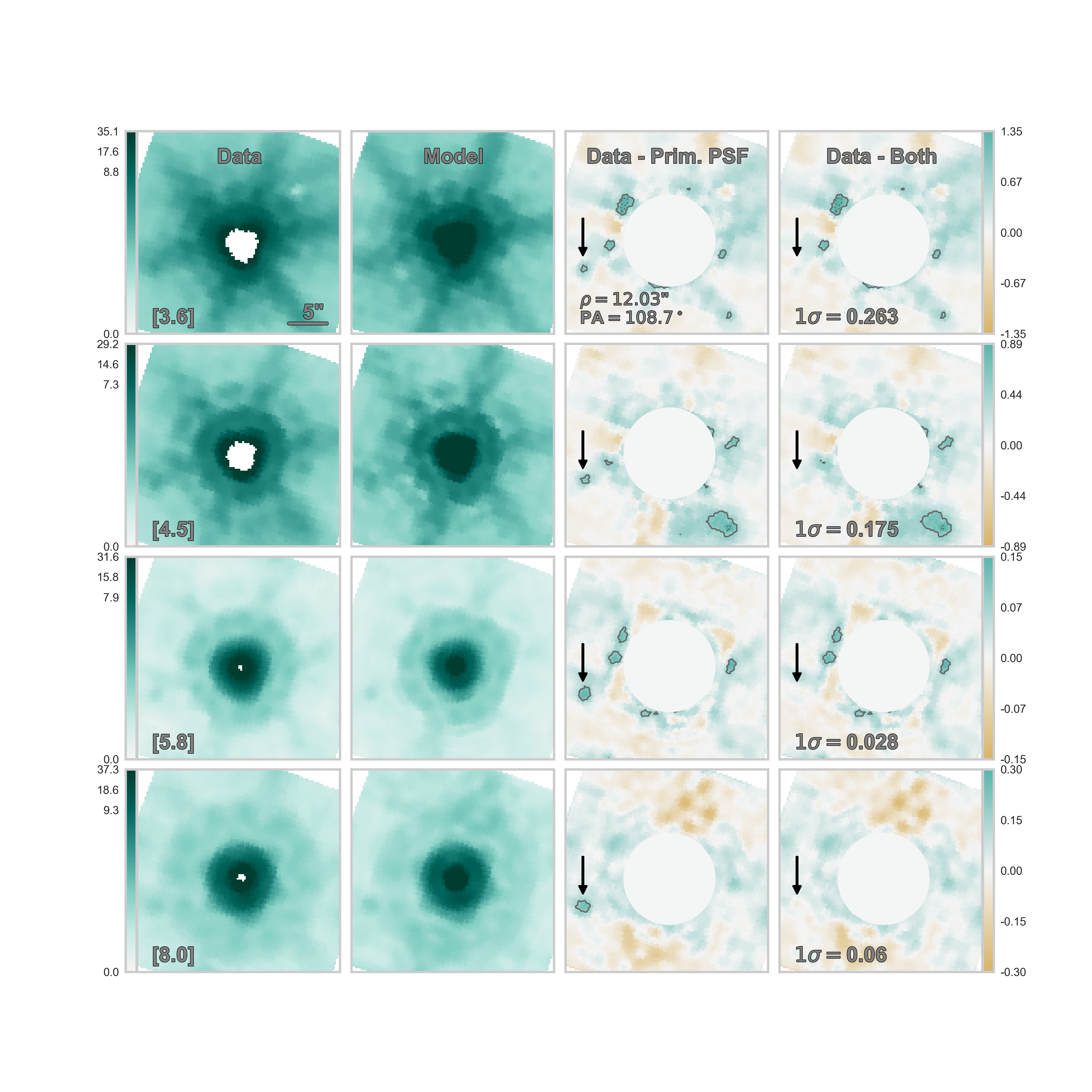}
    \caption{Stacked images of HD 203030 across all four IRAC channels (rows) after its first epoch images have gone through the PSF-fitting pipeline. The images were generated in the same fashion as AB Pic in Figure \ref{fig:abpic}. Columns 1 and 2 show the original IRAC data of HD 203030 and the median two-source PSF model, respectively, displayed with a logarithmic color scale (leftmost color bar). Column 3 shows the residuals left behind after only the primary PSF model is subtracted from the data. Column 4 shows the residuals left behind after the two-source PSF model is subtracted from the data. The standard deviation of the pixel values outside a $\sim$6$\arcsec$ radius from the primary centroid is displayed in the lower left-hand corner of Column 4 in units of DN/s. Both Columns 3 and 4 are displayed in a linear color scale (rightmost color bar) with the area within $\sim$6$\arcsec$ of the primary centroid masked, and 3- and 5-$\sigma$ contours overlaid with solid and dotted lines, respectively.  After subtracting the primary star PSF, a statistically significant positive residual is seen at the expected position of HD 203030 b. This residual disappears after subtracting the best-fit system PSFs and no significant structure is left behind at the location of the companion.}
    \label{fig:hd203030}
\end{figure*}

To test if this color excess might emerge from our reduction procedures, in Figure \ref{fig:hd203030} we show the stacked image output for our pipeline fits of the first epoch of long-exposure IRAC images. A significant positive residual is present at the expected location of HD 203030 b when only the primary PSF is subtracted, and no significant structure remains at that location when both primary and companion PSFs are subtracted, though in Channels 3 and 4, there appears to be a slight over-subtraction. The maximum pixel value at the location of HD 203030 b prior to PSF subtraction is 0.1359 DN/s and 0.2507 DN/s, respectively in those channels. After PSF subtraction, that pixel value is -0.0058 DN/s and -0.0237 DN/s, or 4\% and 9\% of the initial pixel value. If the over-subtraction is uniform for all pixels in an aperture, this would result in a maximum flux overestimation of 0.05 mag and 0.10 mag in Channels 3 and 4, still not enough to explain the differences between our IRAC photometry and those of \cite{miles-paez17}. The assumption of uniform over-subtraction across an aperture is unrealistic though, and we proceed with an empirical approach to better estimate our uncertainties.

The number of resolution elements in the stacked residuals images is about 1220, 780, 470, and 250 for IRAC Channels 1, 2, 3, and 4, respectively. Thus, we would expect 3, 2, 1, and $<$1 spurious 3-$\sigma$ outliers in the residuals images for each channel. For HD 203030, it appears there are more outliers than expected from noise but many are aligned with the PSF structures most visible for the brightest primaries of our sample. We note that in Channel 2, the large lower right-hand residual is a background object ($\pi$=$0.56\pm0.12$ mas; \citealt{gaia18}). The HD 203030 b residual is present in all channels, bolstering confidence in the detections that do not fall upon PSF structure.

To estimate a systematic uncertainty from PSF modeling, we consider the rms of flux values measured within apertures of 1 FWHM radius at the radial separation of the companion around HD 203030 when calculating the detection limit in the PSF-subtracted images. In Channel 3, we find the rms to be 3.26 DN/s which is 19.9\% of the flux measured for HD 203030 b. Similarly in Channel 4, we find the rms of flux values to be 9.25 DN/s, or 24.0\% of the measured flux of the companion. We therefore conclude that there is no clear evidence of systematic errors in our PSF fit, as the overluminosity would be a 4--5-$\sigma$ effect for each of the [5.8] and [8.0] filters, which sample the IRAC PSF in different ways. However, it is unlikely this discrepancy can be resolved without further observations to independently determine its mid-infrared brightness.

\subsubsection{SR 12}
Observations of SR 12 were a part of the \textit{Spitzer} c2d Legacy survey \citep{evans09} that imaged five nearby molecular clouds with the IRAC and MIPS instruments. Various studies \citep[e.g.,][]{cieza07,cieza09,gutermuth09,gunther14,esplin20} have reported IRAC photometry for SR 12 AB from these data, ranging from 8.16--8.27 mag at 3.6 $\mu$m, 8.16--8.25 mag at 4.5 $\mu$m, 7.99--8.12 mag at 5.8 $\mu$m, and 8.03--8.12 mag at 8.0 $\mu$m, with typical uncertainties between 0.02 and 0.06 mag. Our pipeline photometry for SR 12 AB agrees with these previous measurements within the uncertainties in Channels 1, 3, and 4, though our measurement is $\sim$0.06 mag brighter in Channel 2.

The c2d IRAC photometry for SR 12 c is $m_{[3.6]}=13.65\pm0.08$, $m_{[4.5]}=13.60\pm0.03$ mag, $m_{[5.8]}=13.20\pm0.28$ mag, and $m_{[8.0]}=12.50\pm0.37$ \citep{cieza07,alves10,gunther14}. Our Channel 1 and 2 photometry are significantly discrepant ($m_{[3.6]}=13.99\pm0.11$, $m_{[4.5]}=13.18\pm0.07$ mag), but we are able to constrain SR 12 c's Channel 3 and Channel 4 photometry to $m_{[5.8]}=13.09\pm0.15$ mag, and $m_{[8.0]}=12.17\pm0.13$ mag.

SR 12 c has a spectral type of M9-L0 \citep[e.g.,][]{kuzuhara11,bowler14,santamaria18} and thus its photosphere should have a $K_s$--[3.6] color of $\sim$0.6--0.7 mag based on empirical measurements of late M dwarfs \citep[e.g.,][]{patten06,luhman10}. \cite{kuzuhara11} reported ground-based photometry of $K_s=14.57\pm0.03$ mag with their discovery of SR 12 c. Combining this $K_s$-band measurement with our IRAC Channel 1 photometry gives $K_s-[3.6]=0.58\pm0.11$ mag, consistent with a detection of an M9 photosphere. We also measure $[3.6]-[8.0]=1.82\pm0.17$ mag which indicates the companion harbors a disk. \cite{santamaria18} identified numerous emission line accretion tracers in the spectrum of SR 12 c, confirming this disk.

The 4.5 $\mu$m photometry we measure for the companion is the most discrepant from previous studies. No \textit{WISE} photometry has been reported for SR 12 c either, likely due to its crowded environment. Since our photometry of SR 12 AB agrees with previously reported values, any uncertainties would likely come from PSF subtraction. We again consider the rms of flux values measured using apertures with radius equal to 1 FWHM at the radial separation of the companion around SR 12 AB when calculating the detection limit in the PSF-subtracted images. We find at 4.5 $\mu$m the rms to be 1.75 DN/s which is 6.4\% of the flux measured for SR 12 c. Similarly at 5.8 $\mu$m, we find the rms of flux values to be 4.63 DN/s, or 14.2\% of the measured flux of the companion. We conclude here that any large-scale deviation from an M9 photosphere may be outlining the SED of the disk harbored by SR 12 c.

As we mentioned in Section \ref{data_analysis}, SR 12 is $25\arcsec$ away from YLW 13B, a bright and saturated young stellar object. The discrepancies between the c2d companion photometry and ours photometry could be a result of the c2d pipeline's handling of bright neighbors, especially in Channel 1. Conversely, if there is a disk excess in the IRAC bands then it seems likely that we either overestimate the brightness at 4.5 $\mu$m or underestimate the brightness at 5.8 $\mu$m.

\begin{figure*}
    \centering
    \includegraphics[trim=1.0cm 0.25cm 0.25cm 0.0cm,clip,width=1.0\textwidth]{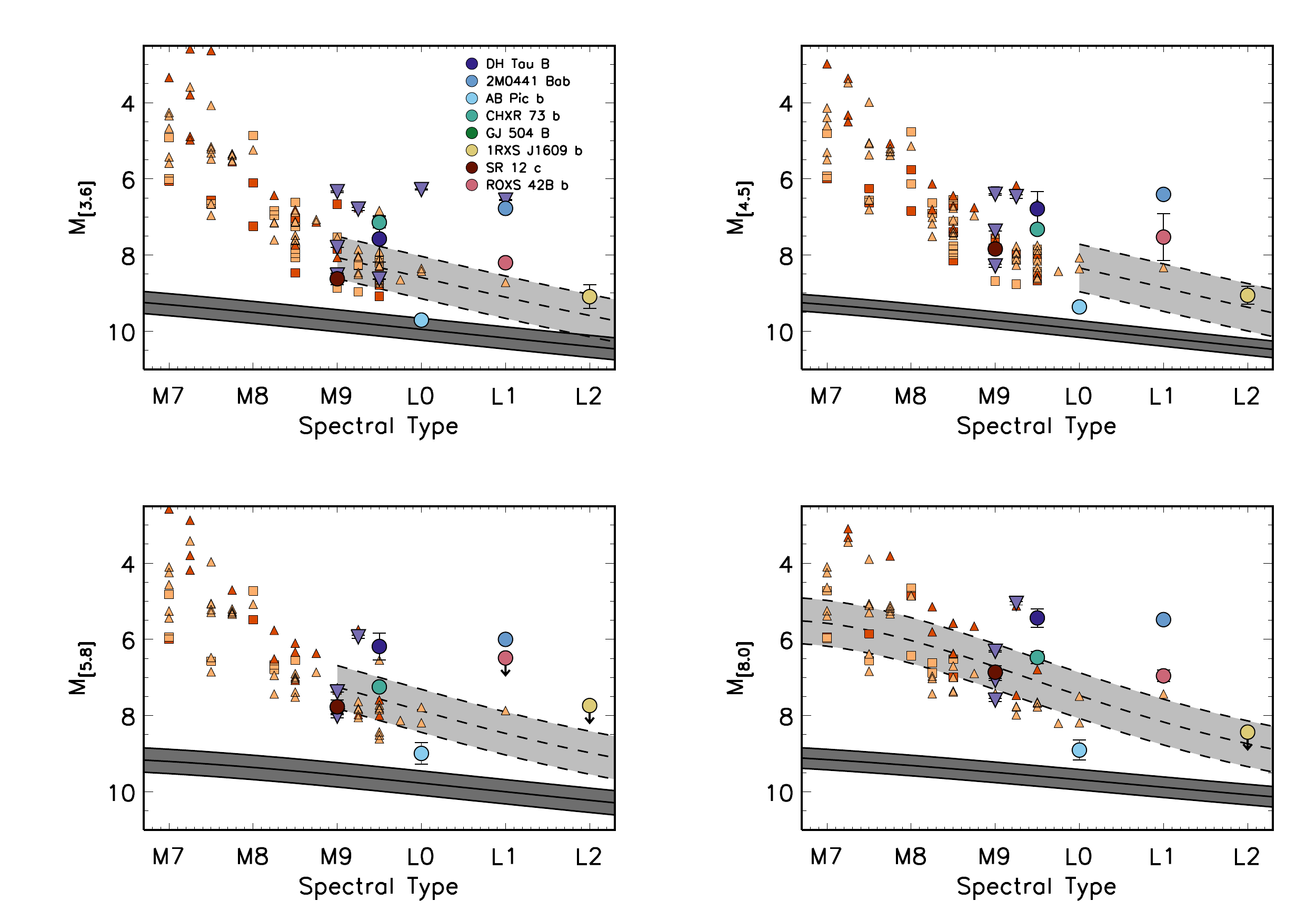}
    \caption{$M_{[3.6]}$ through $M_{[8.0]}$ vs.~spectral type for the $\leq$L2 wide-orbit PMC samples of this work and Paper I (purple upside triangles), in addition to the young Taurus and Upper Sco brown dwarfs (depicted as in Figure \ref{fig:cmd}). Similar to ROXs 42B b, 1RXS J1609 b was not detected in Channel 1 but is shown in the upper lefthand panel as an $L^{\prime}$-band detection from \citet{kraus14}. We also indicate the expected field polynomial sequence of \cite{dupuy12} (solid line; dark gray) and the young ultracool dwarf polynomial sequence from \cite{faherty16} (dashed line; light gray). The young objects sit above the field sequence, but the dynamic range is not high enough in magnitude space to distinguish between disk-bearing and disk-free members.}
    \label{fig:absmag_vs_sptype}
\end{figure*}

\begin{figure}
    \centering
    \includegraphics[trim=1.0cm 0.25cm 0.0cm 0.0cm,clip,width=0.5\textwidth]{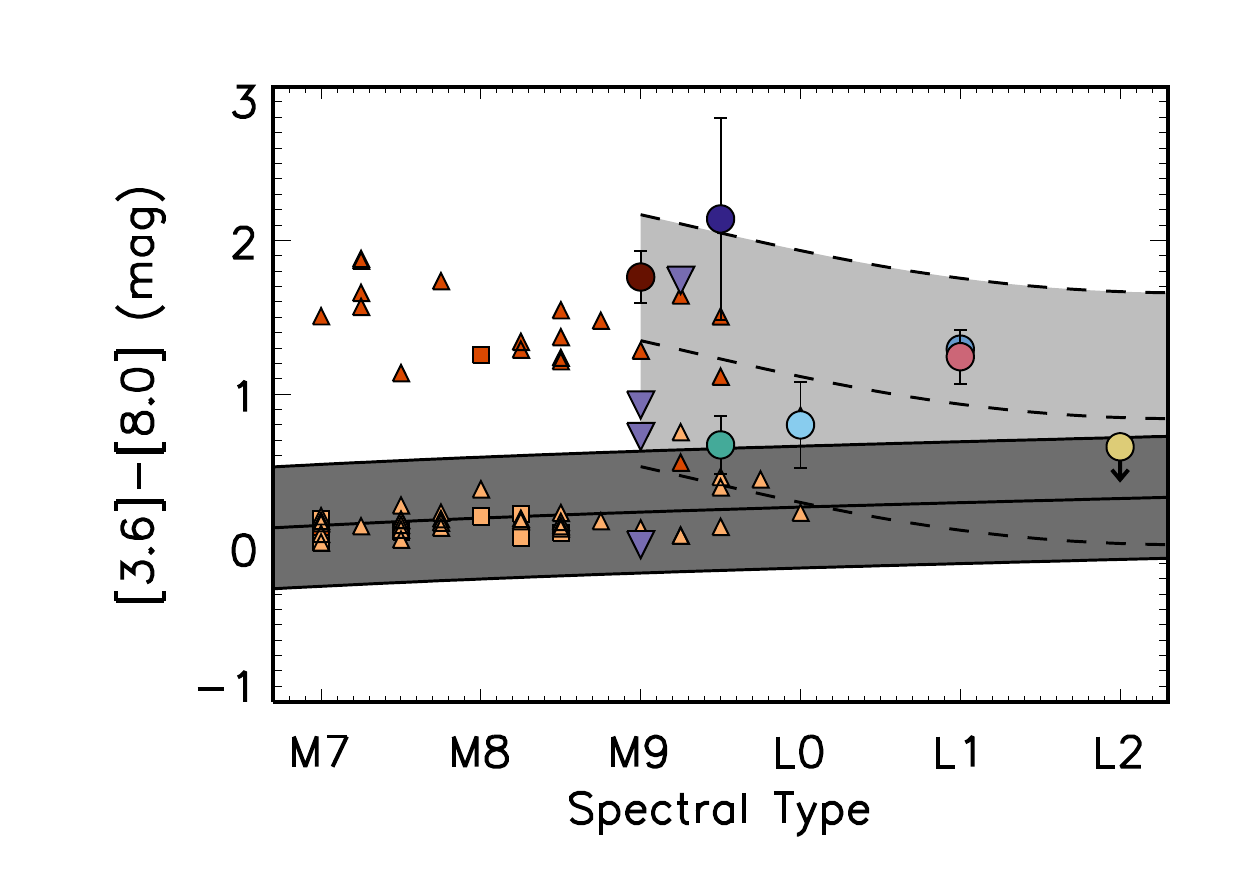}
    \caption{[3.6]--[8.0] color as a function of spectral type for our sample.We include the wide-orbit PMC sample from Paper I, and the young Taurus and Upper Sco brown dwarfs, depicted as in Figures \ref{fig:cmd} and \ref{fig:absmag_vs_sptype}. Also included are field and young moving group (YMG) member polynomial sequences of \cite{dupuy12} (solid line; dark gray) and \cite{faherty16} (dashed line; light gray). The dynamic range is refined enough in color space for the disk-bearing objects to clearly sit above the disk-free objects.}
    \label{fig:i1i4col_vs_sptype}
\end{figure}

\begin{figure*}
    \centering
    \includegraphics[trim=0.90cm 0.25cm 0.50cm 0.5cm,clip,width=0.95\textwidth]{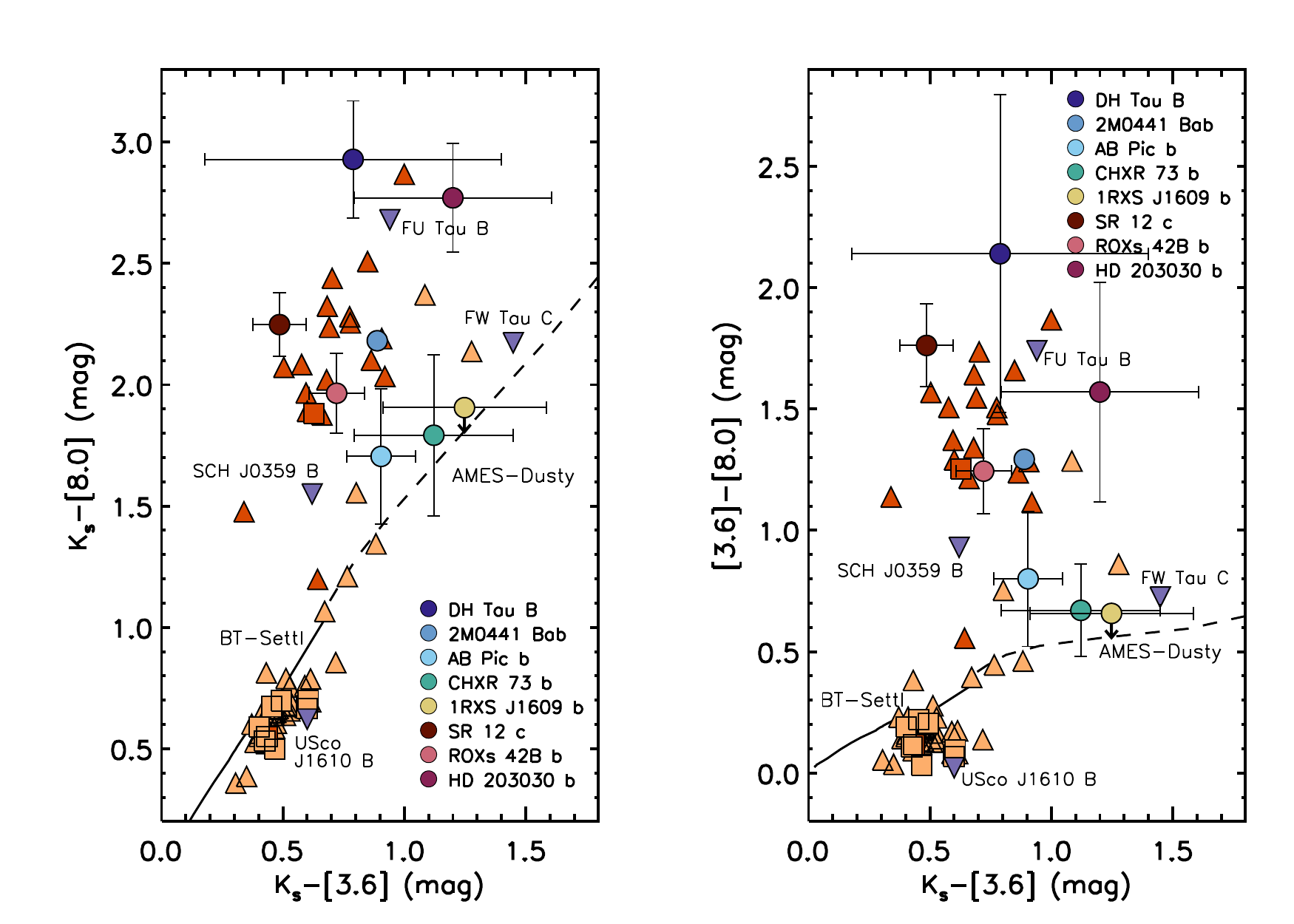}
    \caption{$K_s$--[8.0] vs.~$K_s$--[3.6] color (left panel) and [3.6]--[8.0] vs.~$K_s$--[3.6] color (right panel) for our sample companions, the Paper I wide-orbit PMC sample, and the young Taurus and Upper Sco brown dwarfs, depicted as in Figures \ref{fig:cmd}, \ref{fig:absmag_vs_sptype} and \ref{fig:i1i4col_vs_sptype}. We show both of these color spaces to take advantage of better constrained ground-based $K$-band contrasts for some companions that were marginally detected by our pipeline at [3.6]. We also include the expected color-color sequence of 5 Myr BT-Settl and AMES-Dusty isochrones. The disk-bearing objects are clear outliers in these particular color-color spaces, providing a criterion to say DH Tau B, 2M0441 B, ROXs 42B b, and SR 12 c appear to host disks.}
    \label{fig:i1i4col_vs_ksi1_v1}
\end{figure*}

\section{Discussion}
\label{discussion}
Free-floating young brown dwarfs are observed to follow color-magnitude sequences that are distinct from older brown dwarfs \citep[e.g.,][]{allers10,liu16,faherty16}. In the near-infrared young brown dwarfs are redder in $J-K$ colors, suggesting enhanced dust abundances \citep[e.g.,][]{woitke04,barman11a} or lower surface gravities \citep[e.g.,][]{burrows97,kirkpatrick06,looper08}. Determining whether wide-orbit PMCs also follow the trends previously established for free-floating brown dwarfs into the mid-infrared could point to formation pathway commonalities. Deviations would imply differing formation processes and redder colors could indicate the presence of circum(sub)stellar disks.

\subsection{Absolute Magnitude Trends with Spectral Type}
A star forms with a large radius that subsequently contracts in its pre-main sequence phase, which might result in an observable difference between the luminosities of young stars and substellar objects than those of the field. These objects would also appear brighter in the \textit{Spitzer}/IRAC bands, especially the later spectral types and objects that harbor disks.

Figure \ref{fig:absmag_vs_sptype} shows absolute magnitude-spectral type diagrams plotting $M_{[3.6]}$ through $M_{[8.0]}$ versus spectral type for the detected wide-orbit companions in our sample (filled circles) as well as others from Paper I (FU Tau B, FW Tau C, SCH J0359 B, USco 1610 B), ROXs 12 B, GQ Lup B, and GSC 6214 B (purple upside down triangles). We complement these data with late-M to early-L brown dwarfs from the Taurus and Upper Sco star-forming regions \citep{esplin17,luhman20}. The absolute magnitudes for the individual PMCs were calculated from either the \textit{Gaia} EDR3 or DR2 parallactic measurements \citep{bailer-jones21,bailer-jones18}, or if not available, from the adopted distance to the star-forming region. Individual association members are color-coded red if they are thought to harbor a disk from measured mid-infrared excess, or orange if they are thought to be disk-free. We also indicate the expected field polynomial sequence as determined by \cite{dupuy12} (solid line; dark gray) as well as the young ($\tau<1$ Gyr, $\tau\sim5$--150 Myr; \citealt{faherty16}, \citealt{liu16}) ultracool dwarf polynomial sequence from \cite{faherty16} (dashed line; light gray).

In general, brown dwarfs with spectral types $<$M8 are 1--2 magnitudes brighter than the YMG polynomial sequence, while substantial overlap begins between the YMG sequence and brown dwarfs with spectral types $>$M8. This overluminosity above the field sequence is expected as the young objects have not yet contracted to their final radii. DH Tau B, 2M0441 B, CHXR 73 b, and ROXs 42B b are consistently above the YMG polynomial sequence, as well as FU Tau B. These wide companions orbit host stars that are among the very young regions ($\tau\sim1$--3 Myr; Taurus, Chameleon, Ophiuchus). AB Pic b is the only PMC in our sample that is consistently below the YMG sequence. The high scatter within these sequences suggests that magnitudes alone do not provide a sensitive view of which objects are outliers.

\subsection{Color Trends of Wide-orbit Companions in the Mid-Infrared}
The colors of wide-orbit PMCs provide a more nuanced view of their non-photospheric behavior. The colors of our sample are expected to be close to zero given their range in spectral types, thus objects with non-zero colors are potentially interesting.

In Figure \ref{fig:i1i4col_vs_sptype} we show [3.6]--[8.0] color as a function of spectral type for the same systems as described above for Figure \ref{fig:absmag_vs_sptype}. We again indicate the Taurus (triangles) and Upper Sco (squares) members as disk-bearing (red) or disk-free (orange), and include the field and YMG member polynomial sequences of \cite{dupuy12} (solid line; dark gray) and \cite{faherty16} (dashed line; light gray).

DH Tau B, 2M0441 B, AB Pic b, CHXR 73 b, SR 12 c, and ROXs 42B b are significantly redder than the field polynomial sequence. The young ($\tau\sim2$--10 Myr) Taurus and Upper Sco disk-hosting and disk-free members also readily differentiate themselves in the [3.6]--[8.0] color space.

Interestingly, the disk-bearing members fall right in line with the continuation of the YMG ($\sim$20--120 Myr) dwarf sequence. The detected PMCs of this sample also are consistent with the YMG sequence except for DH Tau B, which is already known to show active accretion. 2M0441 B, SR 12 c, and ROXs 42B b are above the average YMG polynomical sequence color for their spectral type which could be due to the youth of the systems or the presence of circum(sub)stellar disks. There also is the possibility that some YMG members may also harbor circum(sub)stellar disks.

\subsection{Identifying Disk-Hosting PMCs in Color-Color Space}
\label{sec:pmc_disks}
Identifying disk hosts in color-color space removes reliance on spectral type measurements that can be highly uncertain. In Figure \ref{fig:i1i4col_vs_ksi1_v1} we show $K_s$--[8.0] vs.~$K_s$--[3.6] color (left panel) and [3.6]--[8.0] vs.~$K_s$--[3.6] color (right panel) for our PMC sample and the young Taurus and Upper Sco brown dwarfs, depicted as in Figures \ref{fig:absmag_vs_sptype} and \ref{fig:i1i4col_vs_sptype}. We also include the expected color-color sequence of 5 Myr BT-Settl and AMES-Dusty isochrones as a theoretical comparison. We show both of these color spaces to take advantage of better constrained ground-based $K$-band contrasts for some companions that were marginally detected by our pipeline at [3.6].

Five of the wide orbit PMCs in this work (DH Tau B, 2M0441B, ROXs 42B b, SR 12 c, and HD 203030 b), along with two from Paper I (FU Tau B and SCH J0359 B) have colors consistent with young disk-bearing brown dwarfs. However, HD 203030 b is the latest spectral type of our sample (L7.5) thus its position in this parameter space is likely explained by differences in late-L atmospheric characteristics rather than the presence of a circum(sub)stellar disk. Only one object in this combined sample, the more massive USco 1609 B ($M\sim70$ $M_{\mathrm{Jup}}$), falls among the disk-free young brown dwarfs. AB Pic b and CHXR 73 b fall outside of the disk-free locus, along with a few late-type disk-free members, but their locations are consistent with predictions from the AMES-Dusty models \citep{chabrier00b,allard01}. FW Tau C sits furthest from the disk-hosting and disk-free objects, but given the ongoing debate over whether it a more massive object hosting an edge-on disk \citep{wu17a}, it might be expected to have anomalous colors.

\subsection{Disk Fraction of Wide-Orbit PMCs}
\label{sec:disk_fraction}
Determining the presence of circumstellar disks around young star-forming region members has been a useful tool to infer the dominant formation pathway of substellar objects, as well as their planet-forming capabilities. Similarly, identifying and characterizing the disks harbored by PMCs offers a direct avenue to study planet assembly and evolution, as well as potential satellite formation.

Mass-dependent disk evolution has been observed for stars and brown dwarfs in young star-forming regions or associations through the measurement of disk fractions. \cite{luhman10} found the disk fraction for solar-type stars in Taurus ($\tau\sim2$ Myr) to be $\sim$75\% and the disk fraction for lower-mass stars (0.01--0.3 $M_{\odot}$) to be $\sim$45\%. For the older Upper Sco OB association ($\tau\sim10$ Myr), \cite{carpenter06} find $<$1\% of stars more massive than K0 have circumstellar disks while the disk fraction for K0--M5 stars is 19\%. Substantial disk fractions persisting for stars $<$1 $M_{\odot}$ and substellar objects indicate disk dispersal is less efficient and that planet formation timescales are longer. We can now begin to quantify whether these disk frequency trends continue for wide-orbit PMCs. For instance, \cite{bowler17} found $46\%\pm14\%$ of young ($<$15 Myr) substellar ($<$20 $M_{\rm Jup}$) companions have detectable Pa$\beta$ emission, indicating that accretion disks are very common around wide-orbit PMCs. Here, we incorporate our findings into previous disk fraction determinations and explore their global frequency.

Combining the nine PMC systems from this work with three from Paper 1, ten belong to star-forming regions or associations with $\tau<15$ Myr: DH Tau, SCH J0359, FU Tau, FW Tau, 2M0441, AB Pic, CHXR 73, ROXs 42B, 1RXS J1609, and SR 12. Since 2M0441 is an interesting quadruple system comprised of close binary pairs, they should be considered separately and not incorporated into our disk fraction calculation. Thus, six of these companions have disk-like mid-infrared excesses determined from this work, suggesting a disk frequency of $67\%\pm16\%$ for PMCs with $\tau<15$ Myr. The two older PMC systems in our sample, GJ 504 and HD 203030, host companions that do not have disk-like mid-infrared excesses.

Previous PMC disk fraction determinations from \cite{bowler17} and \cite{bryan20} required emission line accretion signatures or UV continuum excess detections to designate a companion as a disk host, potentially underestimating their occurrence rate measurement because of the variability of these signatures or the overall faintness of the disk. Here we combine our PMC sample disk determinations with their findings, updating ROXs 42B b and SR 12 c as disk-bearing, giving a disk fraction of $56\%\pm12\%$. This confirms that PMCs harboring circum(sub)stellar disks is very common at young ages. Even within our $<$15 Myr age bin, hints of PMC disk evolution may be emerging since two of the three companions with no mid-infrared excess from this work had system ages above 5 Myr. Increasing the sample of $>$5 Myr PMC systems with and without circum(sub)stellar disks will ultimately confirm whether the rate at with which they host disks follows that observed of star-forming region members.

\section{Summary}
We have used our MCMC-based PSF formalism to reanalyze \textit{Spitzer}/IRAC images of nine stars known to host faint planetary-mass companions, examining higher contrast systems and closer-in separations than our previous work to measure the mid-infrared photometry of the companions. We report new IRAC photometry for all nine primaries in our sample and eight of the companions, five of which have not been resolved in IRAC images before.

For one of the newly resolved companions, AB Pic b, we use our photometry and the updated system age of 13 Myr \citep{booth2021} to estimate its mass at $M=11\pm1$ $M_\mathrm{Jup}$, placing the companion firmly below the deuterium-burning limit. We also measure an 8.0 $\mu$m excess for ROXs 42B b, a companion not thought to harbor a disk due to a lack of observed emission line accretion signatures. We also confirm mid-infrared excesses from the previously suggested disks around DH Tau B, 2M0441 B, and SR 12 c, and detect likely photospheric emission from four companions that do not show evidence of disks (AB Pic b, CHXR 73 b, 1RXS J1609 b, HD 203030 b).

We find for our sample from Paper I and this work that $67\%\pm16\%$ of young ($<$15 Myr) wide-orbit PMCs harbor disks. Combined with past detections of disk indicators to wide-orbit PMCs, we find a global young disk fraction of $56\%\pm12\%$, signifying that both accreting and non-accreting PMC disks are very common. The increasing likelihood that the disks surrounding wide-orbit PMCs are compact and optically thick, and thus easier to study in the mid-infrared \citep{wu17b}, highlights the importance of leveraging \textit{Spitzer} to motivate future observations of PMC systems in the \textit{JWST} era.

\begin{acknowledgments}
We thank the referee for providing a helpful review that improved the clarity of this paper. R.A.M.~acknowledges support from the Donald D.~Harrington Fellowship and the NASA Earth \& Space Science Fellowship. This work is based on observations made with the \textit{Spitzer Space Telescope}, which is operated by the Jet Propulsion Laboratory, California Institute of Technology under a contract with NASA. This publication makes use of data products from the Two Micron All Sky Survey, which is a joint project of the University of Massachusetts and the Infrared Processing and Analysis Center/California Institute of Technology, funded by the National Aeronautics and Space Administration and the National Science Foundation. This publication makes use of data products from the \textit{Wide-field Infrared Survey Explorer}, which is a joint project of the University of California, Los Angeles, and the Jet Propulsion Laboratory/California Institute of Technology, funded by the National Aeronautics and Space Administration. The Pan-STARRS1 Surveys (PS1) and the PS1 public science archive have been made possible through contributions by the Institute for Astronomy, the University of Hawaii, the Pan-STARRS Project Office, the Max-Planck Society and its participating institutes, the Max-Planck Institute for Astronomy, Heidelberg and the Max-Planck Institute for Extraterrestrial Physics, Garching, The Johns Hopkins University, Durham University, the University of Edinburgh, the Queen’s University Belfast, the Harvard-Smithsonian Center for Astrophysics, the Las Cumbres Observatory Global Telescope Network Incorporated, the National Central University of Taiwan, the Space Telescope Science Institute, the National Aeronautics and Space Administration under grant No.~NNX08AR22G issued through the Planetary Science Division of the NASA Science Mission Directorate, the National Science Foundation grant No.~AST-1238877, the University of Maryland, Eotvos Lorand University (ELTE), the Los Alamos National Laboratory, and the Gordon and Betty Moore Foundation. This work is also based on observations made with the NASA/ESA \textit{Hubble Space Telescope}, and obtained from the Hubble Legacy Archive, which is a collaboration between the Space Telescope Science Institute (STScI/NASA), the Space Telescope European Coordinating Facility (ST-ECF/ESA) and the Canadian Astronomy Data Centre (CADC/NRC/CSA).
\end{acknowledgments}

\clearpage
\bibliographystyle{aasjournal}
\bibliography{swc_ii.bib}

\begin{thebibliography}{}
\expandafter\ifx\csname natexlab\endcsname\relax\def\natexlab#1{#1}\fi
\providecommand{\url}[1]{\href{#1}{#1}}

\bibitem[{{Adame} {et~al.}(2011){Adame}, {Calvet}, {Luhman}, {D'Alessio},
  {Furlan}, {McClure}, {Hartmann}, {Forrest}, \& {Watson}}]{adame11}
{Adame}, L., {Calvet}, N., {Luhman}, K.~L., {et~al.} 2011, \apjl, 726, L3

\bibitem[{{Allard} {et~al.}(2001){Allard}, {Hauschildt}, {Alexander},
  {Tamanai}, \& {Schweitzer}}]{allard01}
{Allard}, F., {Hauschildt}, P.~H., {Alexander}, D.~R., {Tamanai}, A., \&
  {Schweitzer}, A. 2001, \apj, 556, 357

\bibitem[{{Allard} {et~al.}(2012){Allard}, {Homeier}, \& {Freytag}}]{allard12}
{Allard}, F., {Homeier}, D., \& {Freytag}, B. 2012, Philosophical Transactions
  of the Royal Society of London Series A, 370, 2765

\bibitem[{{Allers} {et~al.}(2010){Allers}, {Liu}, {Dupuy}, \&
  {Cushing}}]{allers10}
{Allers}, K.~N., {Liu}, M.~C., {Dupuy}, T.~J., \& {Cushing}, M.~C. 2010, \apj,
  715, 561

\bibitem[{{Alves de Oliveira} {et~al.}(2010){Alves de Oliveira}, {Moraux},
  {Bouvier}, {Bouy}, {Marmo}, \& {Albert}}]{alves10}
{Alves de Oliveira}, C., {Moraux}, E., {Bouvier}, J., {et~al.} 2010, \aap, 515,
  A75

\bibitem[{{Bailer-Jones} {et~al.}(2021){Bailer-Jones}, {Rybizki}, {Fouesneau},
  {Demleitner}, \& {Andrae}}]{bailer-jones21}
{Bailer-Jones}, C.~A.~L., {Rybizki}, J., {Fouesneau}, M., {Demleitner}, M., \&
  {Andrae}, R. 2021, \aj, 161, 147

\bibitem[{{Bailer-Jones} {et~al.}(2018){Bailer-Jones}, {Rybizki}, {Fouesneau},
  {Mantelet}, \& {Andrae}}]{bailer-jones18}
{Bailer-Jones}, C.~A.~L., {Rybizki}, J., {Fouesneau}, M., {Mantelet}, G., \&
  {Andrae}, R. 2018, \aj, 156, 58

\bibitem[{{Bailey} {et~al.}(2014){Bailey}, {Meshkat}, {Reiter}, {Morzinski},
  {Males}, {Su}, {Hinz}, {Kenworthy}, {Stark}, {Mamajek}, {Briguglio}, {Close},
  {Follette}, {Puglisi}, {Rodigas}, {Weinberger}, \& {Xompero}}]{bailey14}
{Bailey}, V., {Meshkat}, T., {Reiter}, M., {et~al.} 2014, \apjl, 780, L4

\bibitem[{{Barman} {et~al.}(2011){Barman}, {Macintosh}, {Konopacky}, \&
  {Marois}}]{barman11a}
{Barman}, T.~S., {Macintosh}, B., {Konopacky}, Q.~M., \& {Marois}, C. 2011,
  \apj, 733, 65

\bibitem[{{Baron} {et~al.}(2018){Baron}, {Artigau}, {Rameau}, {Lafreni{\`e}re},
  {Gagn{\'e}}, {Malo}, {Albert}, {Naud}, {Doyon}, {Janson}, {Delorme}, \&
  {Beichman}}]{baron18}
{Baron}, F., {Artigau}, {\'E}., {Rameau}, J., {et~al.} 2018, \aj, 156, 137

\bibitem[{{Bell} {et~al.}(2015){Bell}, {Mamajek}, \& {Naylor}}]{bell15}
{Bell}, C. P.~M., {Mamajek}, E.~E., \& {Naylor}, T. 2015, \mnras, 454, 593

\bibitem[{{Bonnefoy} {et~al.}(2014){Bonnefoy}, {Chauvin}, {Lagrange}, {Rojo},
  {Allard}, {Pinte}, {Dumas}, \& {Homeier}}]{bonnefoy14}
{Bonnefoy}, M., {Chauvin}, G., {Lagrange}, A.~M., {et~al.} 2014, \aap, 562,
  A127

\bibitem[{{Bonnefoy} {et~al.}(2010){Bonnefoy}, {Chauvin}, {Rojo}, {Allard},
  {Lagrange}, {Homeier}, {Dumas}, \& {Beuzit}}]{bonnefoy10}
{Bonnefoy}, M., {Chauvin}, G., {Rojo}, P., {et~al.} 2010, \aap, 512, A52

\bibitem[{{Booth} {et~al.}(2021){Booth}, {del Burgo}, \&
  {Hambaryan}}]{booth2021}
{Booth}, M., {del Burgo}, C., \& {Hambaryan}, V.~V. 2021, \mnras, 500, 5552

\bibitem[{{Borucki} {et~al.}(2010){Borucki}, {Koch}, {Basri}, {Batalha},
  {Brown}, {Caldwell}, {Caldwell}, {Christensen-Dalsgaard}, {Cochran},
  {DeVore}, {Dunham}, {Dupree}, {Gautier}, {Geary}, {Gilliland}, {Gould},
  {Howell}, {Jenkins}, {Kondo}, {Latham}, {Marcy}, {Meibom}, {Kjeldsen},
  {Lissauer}, {Monet}, {Morrison}, {Sasselov}, {Tarter}, {Boss}, {Brownlee},
  {Owen}, {Buzasi}, {Charbonneau}, {Doyle}, {Fortney}, {Ford}, {Holman},
  {Seager}, {Steffen}, {Welsh}, {Rowe}, {Anderson}, {Buchhave}, {Ciardi},
  {Walkowicz}, {Sherry}, {Horch}, {Isaacson}, {Everett}, {Fischer}, {Torres},
  {Johnson}, {Endl}, {MacQueen}, {Bryson}, {Dotson}, {Haas}, {Kolodziejczak},
  {Van Cleve}, {Chandrasekaran}, {Twicken}, {Quintana}, {Clarke}, {Allen},
  {Li}, {Wu}, {Tenenbaum}, {Verner}, {Bruhweiler}, {Barnes}, \&
  {Prsa}}]{borucki10}
{Borucki}, W.~J., {Koch}, D., {Basri}, G., {et~al.} 2010, Science, 327, 977

\bibitem[{{Bowler}(2016)}]{bowler16}
{Bowler}, B.~P. 2016, \pasp, 128, 102001

\bibitem[{{Bowler} \& {Hillenbrand}(2015)}]{bowler15}
{Bowler}, B.~P., \& {Hillenbrand}, L.~A. 2015, \apjl, 811, L30

\bibitem[{{Bowler} {et~al.}(2014){Bowler}, {Liu}, {Kraus}, \&
  {Mann}}]{bowler14}
{Bowler}, B.~P., {Liu}, M.~C., {Kraus}, A.~L., \& {Mann}, A.~W. 2014, \apj,
  784, 65

\bibitem[{{Bowler} {et~al.}(2017){Bowler}, {Kraus}, {Bryan}, {Knutson},
  {Brogi}, {Rizzuto}, {Mace}, {Vanderburg}, {Liu}, {Hillenbrand}, \&
  {Cieza}}]{bowler17}
{Bowler}, B.~P., {Kraus}, A.~L., {Bryan}, M.~L., {et~al.} 2017, \aj, 154, 165

\bibitem[{{Bryan} {et~al.}(2020){Bryan}, {Ginzburg}, {Chiang}, {Morley},
  {Bowler}, {Xuan}, \& {Knutson}}]{bryan20}
{Bryan}, M.~L., {Ginzburg}, S., {Chiang}, E., {et~al.} 2020, \apj, 905, 37

\bibitem[{{Bulger} {et~al.}(2014){Bulger}, {Patience}, {Ward-Duong}, {Pinte},
  {Bouy}, {M{\'e}nard}, \& {Monin}}]{bulger14}
{Bulger}, J., {Patience}, J., {Ward-Duong}, K., {et~al.} 2014, \aap, 570, A29

\bibitem[{{Burrows} {et~al.}(1997){Burrows}, {Marley}, {Hubbard}, {Lunine},
  {Guillot}, {Saumon}, {Freedman}, {Sudarsky}, \& {Sharp}}]{burrows97}
{Burrows}, A., {Marley}, M., {Hubbard}, W.~B., {et~al.} 1997, \apj, 491, 856

\bibitem[{{Carpenter} {et~al.}(2006){Carpenter}, {Mamajek}, {Hillenbrand}, \&
  {Meyer}}]{carpenter06}
{Carpenter}, J.~M., {Mamajek}, E.~E., {Hillenbrand}, L.~A., \& {Meyer}, M.~R.
  2006, \apjl, 651, L49

\bibitem[{{Chabrier} {et~al.}(2000){Chabrier}, {Baraffe}, {Allard}, \&
  {Hauschildt}}]{chabrier00b}
{Chabrier}, G., {Baraffe}, I., {Allard}, F., \& {Hauschildt}, P. 2000, \apj,
  542, 464

\bibitem[{{Chambers} {et~al.}(2016){Chambers}, {Magnier}, {Metcalfe},
  {Flewelling}, {Huber}, {Waters}, {Denneau}, {Draper}, {Farrow}, {Finkbeiner},
  {Holmberg}, {Koppenhoefer}, {Price}, {Saglia}, {Schlafly}, {Smartt},
  {Sweeney}, {Wainscoat}, {Burgett}, {Grav}, {Heasley}, {Hodapp}, {Jedicke},
  {Kaiser}, {Kudritzki}, {Luppino}, {Lupton}, {Monet}, {Morgan}, {Onaka},
  {Stubbs}, {Tonry}, {Banados}, {Bell}, {Bender}, {Bernard}, {Botticella},
  {Casertano}, {Chastel}, {Chen}, {Chen}, {Cole}, {Deacon}, {Frenk},
  {Fitzsimmons}, {Gezari}, {Goessl}, {Goggia}, {Goldman}, {Grebel}, {Hambly},
  {Hasinger}, {Heavens}, {Heckman}, {Henderson}, {Henning}, {Holman}, {Hopp},
  {Ip}, {Isani}, {Keyes}, {Koekemoer}, {Kotak}, {Long}, {Lucey}, {Liu},
  {Martin}, {McLean}, {Morganson}, {Murphy}, {Nieto-Santisteban}, {Norberg},
  {Peacock}, {Pier}, {Postman}, {Primak}, {Rae}, {Rest}, {Riess}, {Riffeser},
  {Rix}, {Roser}, {Schilbach}, {Schultz}, {Scolnic}, {Szalay}, {Seitz},
  {Shiao}, {Small}, {Smith}, {Soderblom}, {Taylor}, {Thakar}, {Thiel},
  {Thilker}, {Urata}, {Valenti}, {Walter}, {Watters}, {Werner}, {White},
  {Wood-Vasey}, \& {Wyse}}]{chambers16}
{Chambers}, K.~C., {Magnier}, E.~A., {Metcalfe}, N., {et~al.} 2016, ArXiv
  e-prints, arXiv:1612.05560

\bibitem[{{Chauvin} {et~al.}(2005){Chauvin}, {Lagrange}, {Zuckerman}, {Dumas},
  {Mouillet}, {Song}, {Beuzit}, {Lowrance}, \& {Bessell}}]{chauvin05}
{Chauvin}, G., {Lagrange}, A.~M., {Zuckerman}, B., {et~al.} 2005, \aap, 438,
  L29

\bibitem[{{Chinchilla} {et~al.}(2020){Chinchilla}, {B{\'e}jar}, {Lodieu},
  {Gauza}, {Zapatero Osorio}, {Rebolo}, {Garrido}, {Alvarez}, \&
  {Manjavacas}}]{chinchilla20}
{Chinchilla}, P., {B{\'e}jar}, V. J.~S., {Lodieu}, N., {et~al.} 2020, \aap,
  633, A152

\bibitem[{{Cieza} {et~al.}(2007){Cieza}, {Padgett}, {Stapelfeldt}, {Augereau},
  {Harvey}, {Evans}, {Mer{\'{\i}}n}, {Koerner}, {Sargent}, {van Dishoeck},
  {Allen}, {Blake}, {Brooke}, {Chapman}, {Huard}, {Lai}, {Mundy}, {Myers},
  {Spiesman}, \& {Wahhaj}}]{cieza07}
{Cieza}, L., {Padgett}, D.~L., {Stapelfeldt}, K.~R., {et~al.} 2007, \apj, 667,
  308

\bibitem[{{Cieza} {et~al.}(2009){Cieza}, {Padgett}, {Allen}, {McCabe},
  {Brooke}, {Carey}, {Chapman}, {Fukagawa}, {Huard}, {Noriga-Crespo},
  {Peterson}, \& {Rebull}}]{cieza09}
{Cieza}, L.~A., {Padgett}, D.~L., {Allen}, L.~E., {et~al.} 2009, \apjl, 696,
  L84

\bibitem[{{Currie} {et~al.}(2014){Currie}, {Daemgen}, {Debes}, {Lafreniere},
  {Itoh}, {Jayawardhana}, {Ratzka}, \& {Correia}}]{currie14}
{Currie}, T., {Daemgen}, S., {Debes}, J., {et~al.} 2014, \apjl, 780, L30

\bibitem[{{Cutri} {et~al.}(2003){Cutri}, {Skrutskie}, {van Dyk}, {Beichman},
  {Carpenter}, {Chester}, {Cambresy}, {Evans}, {Fowler}, {Gizis}, {Howard},
  {Huchra}, {Jarrett}, {Kopan}, {Kirkpatrick}, {Light}, {Marsh}, {McCallon},
  {Schneider}, {Stiening}, {Sykes}, {Weinberg}, {Wheaton}, {Wheelock}, \&
  {Zacarias}}]{cutri03}
{Cutri}, R.~M., {Skrutskie}, M.~F., {van Dyk}, S., {et~al.} 2003, VizieR Online
  Data Catalog, 2246

\bibitem[{{Cutri} {et~al.}(2021){Cutri}, {Wright}, {Conrow}, {Fowler},
  {Eisenhardt}, {Grillmair}, {Kirkpatrick}, {Masci}, {McCallon}, {Wheelock},
  {Fajardo-Acosta}, {Yan}, {Benford}, {Harbut}, {Jarrett}, {Lake}, {Leisawitz},
  {Ressler}, {Stanford}, {Tsai}, {Liu}, {Helou}, {Mainzer}, {Gettngs},
  {Gonzalez}, {Hoffman}, {Marsh}, {Padgett}, {Skrutskie}, {Beck}, {Papin}, \&
  {Wittman}}]{cutri14}
{Cutri}, R.~M., {Wright}, E.~L., {Conrow}, T., {et~al.} 2021, VizieR Online
  Data Catalog, II/328

\bibitem[{{Deming} {et~al.}(2013){Deming}, {Wilkins}, {McCullough}, {Burrows},
  {Fortney}, {Agol}, {Dobbs-Dixon}, {Madhusudhan}, {Crouzet}, {Desert},
  {Gilliland}, {Haynes}, {Knutson}, {Line}, {Magic}, {Mandell}, {Ranjan},
  {Charbonneau}, {Clampin}, {Seager}, \& {Showman}}]{deming13}
{Deming}, D., {Wilkins}, A., {McCullough}, P., {et~al.} 2013, \apj, 774, 95

\bibitem[{{Dupuy} \& {Liu}(2012)}]{dupuy12}
{Dupuy}, T.~J., \& {Liu}, M.~C. 2012, \apjs, 201, 19

\bibitem[{{Durkan} {et~al.}(2016){Durkan}, {Janson}, \& {Carson}}]{durkan16}
{Durkan}, S., {Janson}, M., \& {Carson}, J.~C. 2016, \apj, 824, 58

\bibitem[{{Epchtein} {et~al.}(1997){Epchtein}, {de Batz}, {Capoani},
  {Chevallier}, {Copet}, {Fouqu{\'e}}, {Lacombe}, {Le Bertre}, {Pau}, {Rouan},
  {Ruphy}, {Simon}, {Tiph{\`e}ne}, {Burton}, {Bertin}, {Deul}, {Habing},
  {Borsenberger}, {Dennefeld}, {Guglielmo}, {Loup}, {Mamon}, {Ng}, {Omont},
  {Provost}, {Renault}, {Tanguy}, {Kimeswenger}, {Kienel}, {Garzon}, {Persi},
  {Ferrari-Toniolo}, {Robin}, {Paturel}, {Vauglin}, {Forveille}, {Delfosse},
  {Hron}, {Schultheis}, {Appenzeller}, {Wagner}, {Balazs}, {Holl},
  {L{\'e}pine}, {Boscolo}, {Picazzio}, {Duc}, \& {Mennessier}}]{epchtein97}
{Epchtein}, N., {de Batz}, B., {Capoani}, L., {et~al.} 1997, The Messenger, 87,
  27

\bibitem[{{Esplin} \& {Luhman}(2017)}]{esplin17}
{Esplin}, T.~L., \& {Luhman}, K.~L. 2017, \aj, 154, 134

\bibitem[{{Esplin} \& {Luhman}(2020)}]{esplin20}
---. 2020, \aj, 159, 282

\bibitem[{{Esplin} {et~al.}(2014){Esplin}, {Luhman}, \& {Mamajek}}]{esplin14}
{Esplin}, T.~L., {Luhman}, K.~L., \& {Mamajek}, E.~E. 2014, \apj, 784, 126

\bibitem[{{Evans} {et~al.}(2009){Evans}, {Dunham}, {J{\o}rgensen}, {Enoch},
  {Mer{\'{\i}}n}, {van Dishoeck}, {Alcal{\'a}}, {Myers}, {Stapelfeldt},
  {Huard}, {Allen}, {Harvey}, {van Kempen}, {Blake}, {Koerner}, {Mundy},
  {Padgett}, \& {Sargent}}]{evans09}
{Evans}, II, N.~J., {Dunham}, M.~M., {J{\o}rgensen}, J.~K., {et~al.} 2009,
  \apjs, 181, 321

\bibitem[{{Faherty} {et~al.}(2016){Faherty}, {Riedel}, {Cruz}, {Gagne},
  {Filippazzo}, {Lambrides}, {Fica}, {Weinberger}, {Thorstensen}, {Tinney},
  {Baldassare}, {Lemonier}, \& {Rice}}]{faherty16}
{Faherty}, J.~K., {Riedel}, A.~R., {Cruz}, K.~L., {et~al.} 2016, \apjs, 225, 10

\bibitem[{{Fazio} {et~al.}(2004){Fazio}, {Hora}, {Allen}, {Ashby}, {Barmby},
  {Deutsch}, {Huang}, {Kleiner}, {Marengo}, {Megeath}, {Melnick}, {Pahre},
  {Patten}, {Polizotti}, {Smith}, {Taylor}, {Wang}, {Willner}, {Hoffmann},
  {Pipher}, {Forrest}, {McMurty}, {McCreight}, {McKelvey}, {McMurray}, {Koch},
  {Moseley}, {Arendt}, {Mentzell}, {Marx}, {Losch}, {Mayman}, {Eichhorn},
  {Krebs}, {Jhabvala}, {Gezari}, {Fixsen}, {Flores}, {Shakoorzadeh}, {Jungo},
  {Hakun}, {Workman}, {Karpati}, {Kichak}, {Whitley}, {Mann}, {Tollestrup},
  {Eisenhardt}, {Stern}, {Gorjian}, {Bhattacharya}, {Carey}, {Nelson},
  {Glaccum}, {Lacy}, {Lowrance}, {Laine}, {Reach}, {Stauffer}, {Surace},
  {Wilson}, {Wright}, {Hoffman}, {Domingo}, \& {Cohen}}]{fazio04}
{Fazio}, G.~G., {Hora}, J.~L., {Allen}, L.~E., {et~al.} 2004, \apjs, 154, 10

\bibitem[{{Filippazzo} {et~al.}(2015){Filippazzo}, {Rice}, {Faherty}, {Cruz},
  {Van Gordon}, \& {Looper}}]{filippazzo15}
{Filippazzo}, J.~C., {Rice}, E.~L., {Faherty}, J., {et~al.} 2015, \apj, 810,
  158

\bibitem[{{Fitzpatrick}(1999)}]{fitzpatrick99}
{Fitzpatrick}, E.~L. 1999, \pasp, 111, 63

\bibitem[{{Gaia Collaboration} {et~al.}(2018){Gaia Collaboration}, {Brown},
  {Vallenari}, {Prusti}, {de Bruijne}, {Babusiaux}, {Bailer-Jones}, {Biermann},
  {Evans}, {Eyer}, \& et~al.}]{gaia18}
{Gaia Collaboration}, {Brown}, A.~G.~A., {Vallenari}, A., {et~al.} 2018, \aap,
  616, A1

\bibitem[{{G{\"u}nther} {et~al.}(2014){G{\"u}nther}, {Cody}, {Covey},
  {Hillenbrand}, {Plavchan}, {Poppenhaeger}, {Rebull}, {Stauffer}, {Wolk},
  {Allen}, {Bayo}, {Gutermuth}, {Hora}, {Meng}, {Morales-Calder{\'o}n},
  {Parks}, \& {Song}}]{gunther14}
{G{\"u}nther}, H.~M., {Cody}, A.~M., {Covey}, K.~R., {et~al.} 2014, \aj, 148,
  122

\bibitem[{{Gutermuth} {et~al.}(2009){Gutermuth}, {Megeath}, {Myers}, {Allen},
  {Pipher}, \& {Fazio}}]{gutermuth09}
{Gutermuth}, R.~A., {Megeath}, S.~T., {Myers}, P.~C., {et~al.} 2009, \apjs,
  184, 18

\bibitem[{{Haffert} {et~al.}(2019){Haffert}, {Bohn}, {de Boer}, {Snellen},
  {Brinchmann}, {Girard}, {Keller}, \& {Bacon}}]{haffert19}
{Haffert}, S.~Y., {Bohn}, A.~J., {de Boer}, J., {et~al.} 2019, Nature
  Astronomy, 3, 749

\bibitem[{{Herbig}(1977)}]{herbig77}
{Herbig}, G.~H. 1977, \apj, 214, 747

\bibitem[{{Hoffman}(2005)}]{hoffman05}
{Hoffman}, W.~F. 2005, in Technical Report

\bibitem[{{H{\o}g} {et~al.}(2000){H{\o}g}, {Fabricius}, {Makarov}, {Urban},
  {Corbin}, {Wycoff}, {Bastian}, {Schwekendiek}, \& {Wicenec}}]{hog00}
{H{\o}g}, E., {Fabricius}, C., {Makarov}, V.~V., {et~al.} 2000, \aap, 355, L27

\bibitem[{{Ireland} {et~al.}(2011){Ireland}, {Kraus}, {Martinache}, {Law}, \&
  {Hillenbrand}}]{ireland11}
{Ireland}, M.~J., {Kraus}, A., {Martinache}, F., {Law}, N., \& {Hillenbrand},
  L.~A. 2011, \apj, 726, 113

\bibitem[{{Itoh} {et~al.}(2005){Itoh}, {Hayashi}, {Tamura}, {Tsuji}, {Oasa},
  {Fukagawa}, {Hayashi}, {Naoi}, {Ishii}, {Mayama}, {Morino}, {Yamashita},
  {Pyo}, {Nishikawa}, {Usuda}, {Murakawa}, {Suto}, {Oya}, {Takato}, {Ando},
  {Miyama}, {Kobayashi}, \& {Kaifu}}]{itoh05}
{Itoh}, Y., {Hayashi}, M., {Tamura}, M., {et~al.} 2005, \apj, 620, 984

\bibitem[{{Janson} {et~al.}(2015){Janson}, {Quanz}, {Carson}, {Thalmann},
  {Lafreni{\`e}re}, \& {Amara}}]{janson15}
{Janson}, M., {Quanz}, S.~P., {Carson}, J.~C., {et~al.} 2015, \aap, 574, A120

\bibitem[{{Kirkpatrick} {et~al.}(2006){Kirkpatrick}, {Barman}, {Burgasser},
  {McGovern}, {McLean}, {Tinney}, \& {Lowrance}}]{kirkpatrick06}
{Kirkpatrick}, J.~D., {Barman}, T.~S., {Burgasser}, A.~J., {et~al.} 2006, \apj,
  639, 1120

\bibitem[{{Konopacky} {et~al.}(2013){Konopacky}, {Barman}, {Macintosh}, \&
  {Marois}}]{konopacky13}
{Konopacky}, Q.~M., {Barman}, T.~S., {Macintosh}, B.~A., \& {Marois}, C. 2013,
  Science, 339, 1398

\bibitem[{{Kraus} \& {Hillenbrand}(2009)}]{kraus09b}
{Kraus}, A.~L., \& {Hillenbrand}, L.~A. 2009, \apj, 704, 531

\bibitem[{{Kraus} {et~al.}(2014{\natexlab{a}}){Kraus}, {Ireland}, {Cieza},
  {Hinkley}, {Dupuy}, {Bowler}, \& {Liu}}]{kraus14}
{Kraus}, A.~L., {Ireland}, M.~J., {Cieza}, L.~A., {et~al.} 2014{\natexlab{a}},
  \apj, 781, 20

\bibitem[{{Kraus} {et~al.}(2014{\natexlab{b}}){Kraus}, {Shkolnik}, {Allers}, \&
  {Liu}}]{kraus14b}
{Kraus}, A.~L., {Shkolnik}, E.~L., {Allers}, K.~N., \& {Liu}, M.~C.
  2014{\natexlab{b}}, \aj, 147, 146

\bibitem[{{Kreidberg} {et~al.}(2014){Kreidberg}, {Bean}, {D{\'e}sert},
  {Benneke}, {Deming}, {Stevenson}, {Seager}, {Berta-Thompson}, {Seifahrt}, \&
  {Homeier}}]{kreidberg14}
{Kreidberg}, L., {Bean}, J.~L., {D{\'e}sert}, J.-M., {et~al.} 2014, \nat, 505,
  69

\bibitem[{{Kuzuhara} {et~al.}(2011){Kuzuhara}, {Tamura}, {Ishii}, {Kudo},
  {Nishiyama}, \& {Kandori}}]{kuzuhara11}
{Kuzuhara}, M., {Tamura}, M., {Ishii}, M., {et~al.} 2011, \aj, 141, 119

\bibitem[{{Kuzuhara} {et~al.}(2013){Kuzuhara}, {Tamura}, {Kudo}, {Janson},
  {Kandori}, {Brandt}, {Thalmann}, {Spiegel}, {Biller}, {Carson}, {Hori},
  {Suzuki}, {Burrows}, {Henning}, {Turner}, {McElwain}, {Moro-Mart{\'\i}n},
  {Suenaga}, {Takahashi}, {Kwon}, {Lucas}, {Abe}, {Brandner}, {Egner}, {Feldt},
  {Fujiwara}, {Goto}, {Grady}, {Guyon}, {Hashimoto}, {Hayano}, {Hayashi},
  {Hayashi}, {Hodapp}, {Ishii}, {Iye}, {Knapp}, {Matsuo}, {Mayama}, {Miyama},
  {Morino}, {Nishikawa}, {Nishimura}, {Kotani}, {Kusakabe}, {Pyo}, {Serabyn},
  {Suto}, {Takami}, {Takato}, {Terada}, {Tomono}, {Watanabe}, {Wisniewski},
  {Yamada}, {Takami}, \& {Usuda}}]{kuzuhara13}
{Kuzuhara}, M., {Tamura}, M., {Kudo}, T., {et~al.} 2013, \apj, 774, 11

\bibitem[{{Lachapelle} {et~al.}(2015){Lachapelle}, {Lafreni{\`e}re},
  {Gagn{\'e}}, {Jayawardhana}, {Janson}, {Helling}, \& {Witte}}]{lachapelle15}
{Lachapelle}, F.-R., {Lafreni{\`e}re}, D., {Gagn{\'e}}, J., {et~al.} 2015,
  \apj, 802, 61

\bibitem[{{Lafreni{\`e}re} {et~al.}(2008){Lafreni{\`e}re}, {Jayawardhana}, \&
  {van Kerkwijk}}]{lafreniere08b}
{Lafreni{\`e}re}, D., {Jayawardhana}, R., \& {van Kerkwijk}, M.~H. 2008, \apjl,
  689, L153

\bibitem[{{Lafreni{\`e}re} {et~al.}(2007){Lafreni{\`e}re}, {Marois}, {Doyon},
  {Nadeau}, \& {Artigau}}]{lafreniere07}
{Lafreni{\`e}re}, D., {Marois}, C., {Doyon}, R., {Nadeau}, D., \& {Artigau},
  {\'E}. 2007, \apj, 660, 770

\bibitem[{{Leggett} {et~al.}(2008){Leggett}, {Saumon}, {Albert}, {Cushing},
  {Liu}, {Luhman}, {Marley}, {Kirkpatrick}, {Roellig}, \& {Allers}}]{leggett08}
{Leggett}, S.~K., {Saumon}, D., {Albert}, L., {et~al.} 2008, \apj, 682, 1256

\bibitem[{{Liu} {et~al.}(2016){Liu}, {Dupuy}, \& {Allers}}]{liu16}
{Liu}, M.~C., {Dupuy}, T.~J., \& {Allers}, K.~N. 2016, \apj, 833, 96

\bibitem[{{Looper} {et~al.}(2008){Looper}, {Kirkpatrick}, {Cutri}, {Barman},
  {Burgasser}, {Cushing}, {Roellig}, {McGovern}, {McLean}, {Rice}, {Swift}, \&
  {Schurr}}]{looper08}
{Looper}, D.~L., {Kirkpatrick}, J.~D., {Cutri}, R.~M., {et~al.} 2008, \apj,
  686, 528

\bibitem[{{Luhman} {et~al.}(2010){Luhman}, {Allen}, {Espaillat}, {Hartmann}, \&
  {Calvet}}]{luhman10}
{Luhman}, K.~L., {Allen}, P.~R., {Espaillat}, C., {Hartmann}, L., \& {Calvet},
  N. 2010, \apjs, 186, 111

\bibitem[{{Luhman} \& {Esplin}(2020)}]{luhman20}
{Luhman}, K.~L., \& {Esplin}, T.~L. 2020, \aj, 160, 44

\bibitem[{{Luhman} {et~al.}(2006){Luhman}, {Wilson}, {Brandner}, {Skrutskie},
  {Nelson}, {Smith}, {Peterson}, {Cushing}, \& {Young}}]{luhman06c}
{Luhman}, K.~L., {Wilson}, J.~C., {Brandner}, W., {et~al.} 2006, \apj, 649, 894

\bibitem[{{MacGregor} {et~al.}(2017){MacGregor}, {Wilner}, {Czekala},
  {Andrews}, {Dai}, {Herczeg}, {Kratter}, {Kraus}, {Ricci}, \&
  {Testi}}]{macgregor17}
{MacGregor}, M.~A., {Wilner}, D.~J., {Czekala}, I., {et~al.} 2017, \apj, 835,
  17

\bibitem[{{Marengo} {et~al.}(2006){Marengo}, {Megeath}, {Fazio}, {Stapelfeldt},
  {Werner}, \& {Backman}}]{marengo06}
{Marengo}, M., {Megeath}, S.~T., {Fazio}, G.~G., {et~al.} 2006, \apj, 647, 1437

\bibitem[{{Marocco} {et~al.}(2020){Marocco}, {Eisenhardt}, {Fowler},
  {Kirkpatrick}, {Meisner}, {Schlafly}, {Stanford}, {Garcia}, {Caselden},
  {Cushing}, {Cutri}, {Faherty}, {Gelino}, {Gonzalez}, {Jarrett}, {Koontz},
  {Mainzer}, {Marchese}, {Mobasher}, {Schlegel}, {Stern}, {Teplitz}, \& {Wright
  E.~L. (The Catwise Team)}}]{marocco20}
{Marocco}, F., {Eisenhardt}, P.~R.~M., {Fowler}, J.~W., {et~al.} 2020, VizieR
  Online Data Catalog, II/365

\bibitem[{{Martinez} \& {Kraus}(2019)}]{martinez19}
{Martinez}, R.~A., \& {Kraus}, A.~L. 2019, \aj, 158, 134

\bibitem[{{Metchev} \& {Hillenbrand}(2006)}]{metchev06}
{Metchev}, S.~A., \& {Hillenbrand}, L.~A. 2006, \apj, 651, 1166

\bibitem[{{Meyer} {et~al.}(1997){Meyer}, {Beckwith}, {Herbst}, \&
  {Robberto}}]{meyer97}
{Meyer}, M.~R., {Beckwith}, S.~V.~W., {Herbst}, T.~M., \& {Robberto}, M. 1997,
  \apjl, 489, L173

\bibitem[{{Miles-P{\'a}ez} {et~al.}(2017){Miles-P{\'a}ez}, {Metchev}, {Luhman},
  {Marengo}, \& {Hulsebus}}]{miles-paez17}
{Miles-P{\'a}ez}, P.~A., {Metchev}, S., {Luhman}, K.~L., {Marengo}, M., \&
  {Hulsebus}, A. 2017, \aj, 154, 262

\bibitem[{{Miret-Roig} {et~al.}(2018){Miret-Roig}, {Antoja},
  {Romero-G{\'o}mez}, \& {Figueras}}]{miret-roig18}
{Miret-Roig}, N., {Antoja}, T., {Romero-G{\'o}mez}, M., \& {Figueras}, F. 2018,
  \aap, 615, A51

\bibitem[{{Patience} {et~al.}(2010){Patience}, {King}, {de Rosa}, \&
  {Marois}}]{patience10}
{Patience}, J., {King}, R.~R., {de Rosa}, R.~J., \& {Marois}, C. 2010, \aap,
  517, A76

\bibitem[{{Patten} {et~al.}(2006){Patten}, {Stauffer}, {Burrows}, {Marengo},
  {Hora}, {Luhman}, {Sonnett}, {Henry}, {Raghavan}, {Megeath}, {Liebert}, \&
  {Fazio}}]{patten06}
{Patten}, B.~M., {Stauffer}, J.~R., {Burrows}, A., {et~al.} 2006, \apj, 651,
  502

\bibitem[{{P{\'e}rez} {et~al.}(2019){P{\'e}rez}, {Marino}, {Casassus},
  {Baruteau}, {Zurlo}, {Flores}, \& {Chauvin}}]{perez19}
{P{\'e}rez}, S., {Marino}, S., {Casassus}, S., {et~al.} 2019, \mnras, 488, 1005

\bibitem[{{Petrus} {et~al.}(2020){Petrus}, {Bonnefoy}, {Chauvin}, {Charnay},
  {Marleau}, {Gratton}, {Lagrange}, {Rameau}, {Mordasini}, {Nowak}, {Delorme},
  {Boccaletti}, {Carlotti}, {Houll{\'e}}, {Vigan}, {Allard}, {Desidera},
  {D'Orazi}, {Hoeijmakers}, {Wyttenbach}, \& {Lavie}}]{petrus20}
{Petrus}, S., {Bonnefoy}, M., {Chauvin}, G., {et~al.} 2020, arXiv e-prints,
  arXiv:2012.02798

\bibitem[{{Rameau} {et~al.}(2013){Rameau}, {Chauvin}, {Lagrange}, {Klahr},
  {Bonnefoy}, {Mordasini}, {Bonavita}, {Desidera}, {Dumas}, \&
  {Girard}}]{rameau13}
{Rameau}, J., {Chauvin}, G., {Lagrange}, A.~M., {et~al.} 2013, \aap, 553, A60

\bibitem[{{Santamar{\'\i}a-Miranda} {et~al.}(2018){Santamar{\'\i}a-Miranda},
  {C{\'a}ceres}, {Schreiber}, {Hardy}, {Bayo}, {Parsons}, {Gromadzki}, \&
  {Aguayo Villegas}}]{santamaria18}
{Santamar{\'\i}a-Miranda}, A., {C{\'a}ceres}, C., {Schreiber}, M.~R., {et~al.}
  2018, \mnras, 475, 2994

\bibitem[{{Skemer} {et~al.}(2016){Skemer}, {Morley}, {Zimmerman}, {Skrutskie},
  {Leisenring}, {Buenzli}, {Bonnefoy}, {Bailey}, {Hinz}, {Defr{\'e}re},
  {Esposito}, {Apai}, {Biller}, {Brandner}, {Close}, {Crepp}, {De Rosa},
  {Desidera}, {Eisner}, {Fortney}, {Freedman}, {Henning}, {Hofmann},
  {Kopytova}, {Lupu}, {Maire}, {Males}, {Marley}, {Morzinski}, {Oza},
  {Patience}, {Rajan}, {Rieke}, {Schertl}, {Schlieder}, {Stone}, {Su}, {Vaz},
  {Visscher}, {Ward-Duong}, {Weigelt}, \& {Woodward}}]{skemer16}
{Skemer}, A.~J., {Morley}, C.~V., {Zimmerman}, N.~T., {et~al.} 2016, \apj, 817,
  166

\bibitem[{{Song} {et~al.}(2003){Song}, {Zuckerman}, \& {Bessell}}]{song03}
{Song}, I., {Zuckerman}, B., \& {Bessell}, M.~S. 2003, \apj, 599, 342

\bibitem[{{Todorov} {et~al.}(2010){Todorov}, {Luhman}, \& {McLeod}}]{todorov10}
{Todorov}, K., {Luhman}, K.~L., \& {McLeod}, K.~K. 2010, \apjl, 714, L84

\bibitem[{{Todorov} {et~al.}(2014){Todorov}, {Luhman}, {Konopacky}, {McLeod},
  {Apai}, {Ghez}, {Pascucci}, \& {Robberto}}]{todorov14}
{Todorov}, K.~O., {Luhman}, K.~L., {Konopacky}, Q.~M., {et~al.} 2014, \apj,
  788, 40

\bibitem[{{Torres} {et~al.}(2008){Torres}, {Quast}, {Melo}, \&
  {Sterzik}}]{torres08}
{Torres}, C.~A.~O., {Quast}, G.~R., {Melo}, C.~H.~F., \& {Sterzik}, M.~F. 2008,
  {Young Nearby Loose Associations}, ed. B.~{Reipurth}, Vol.~5, 757

\bibitem[{{Valenti} {et~al.}(1993){Valenti}, {Basri}, \& {Johns}}]{valenti93}
{Valenti}, J.~A., {Basri}, G., \& {Johns}, C.~M. 1993, \aj, 106, 2024

\bibitem[{{van Belle} \& {von Braun}(2009)}]{vanBelle09}
{van Belle}, G.~T., \& {von Braun}, K. 2009, \apj, 694, 1085

\bibitem[{{Wakeford} {et~al.}(2017){Wakeford}, {Sing}, {Kataria}, {Deming},
  {Nikolov}, {Lopez}, {Tremblin}, {Amundsen}, {Lewis}, {Mandell}, {Fortney},
  {Knutson}, {Benneke}, \& {Evans}}]{wakeford17}
{Wakeford}, H.~R., {Sing}, D.~K., {Kataria}, T., {et~al.} 2017, Science, 356,
  628

\bibitem[{{Watson} {et~al.}(2009){Watson}, {Leisenring}, {Furlan}, {Bohac},
  {Sargent}, {Forrest}, {Calvet}, {Hartmann}, {Nordhaus}, {Green}, {Kim},
  {Sloan}, {Chen}, {Keller}, {d'Alessio}, {Najita}, {Uchida}, \&
  {Houck}}]{watson09}
{Watson}, D.~M., {Leisenring}, J.~M., {Furlan}, E., {et~al.} 2009, \apjs, 180,
  84

\bibitem[{{Werner} {et~al.}(2004){Werner}, {Roellig}, {Low}, {Rieke}, {Rieke},
  {Hoffmann}, {Young}, {Houck}, {Brandl}, {Fazio}, {Hora}, {Gehrz}, {Helou},
  {Soifer}, {Stauffer}, {Keene}, {Eisenhardt}, {Gallagher}, {Gautier}, {Irace},
  {Lawrence}, {Simmons}, {Van Cleve}, {Jura}, {Wright}, \&
  {Cruikshank}}]{werner04}
{Werner}, M.~W., {Roellig}, T.~L., {Low}, F.~J., {et~al.} 2004, \apjs, 154, 1

\bibitem[{{Woitke} \& {Helling}(2004)}]{woitke04}
{Woitke}, P., \& {Helling}, C. 2004, \aap, 414, 335

\bibitem[{{Wolff} {et~al.}(2017){Wolff}, {M{\'e}nard}, {Caceres},
  {Lef{\`e}vre}, {Bonnefoy}, {C{\'a}novas}, {Maret}, {Pinte}, {Schreiber}, \&
  {van der Plas}}]{wolff17}
{Wolff}, S.~G., {M{\'e}nard}, F., {Caceres}, C., {et~al.} 2017, \aj, 154, 26

\bibitem[{{Wu} {et~al.}(2017){Wu}, {Close}, {Eisner}, \& {Sheehan}}]{wu17b}
{Wu}, Y.-L., {Close}, L.~M., {Eisner}, J.~A., \& {Sheehan}, P.~D. 2017, \aj,
  154, 234

\bibitem[{{Wu} \& {Sheehan}(2017)}]{wu17a}
{Wu}, Y.-L., \& {Sheehan}, P.~D. 2017, \apjl, 846, L26

\bibitem[{{Wu} {et~al.}(2015){Wu}, {Close}, {Males}, {Barman}, {Morzinski},
  {Follette}, {Bailey}, {Rodigas}, {Hinz}, {Puglisi}, {Xompero}, \&
  {Briguglio}}]{wu15}
{Wu}, Y.-L., {Close}, L.~M., {Males}, J.~R., {et~al.} 2015, \apjl, 807, L13

\bibitem[{{Wu} {et~al.}(2020){Wu}, {Bowler}, {Sheehan}, {Andrews}, {Herczeg},
  {Kraus}, {Ricci}, {Wilner}, \& {Zhu}}]{wu20}
{Wu}, Y.-L., {Bowler}, B.~P., {Sheehan}, P.~D., {et~al.} 2020, \aj, 159, 229

\bibitem[{Zhou {et~al.}(2014)Zhou, Herczeg, Kraus, Metchev, \& Cruz}]{zhou14}
Zhou, Y., Herczeg, G.~J., Kraus, A.~L., Metchev, S., \& Cruz, K.~L. 2014, The
  Astrophysical Journal Letters, 783, L17

\end{thebibliography}

\clearpage

\startlongtable
\begin{deluxetable*}{cccc}
\tablewidth{1.0\textwidth}
\tablecaption{Photometry for Sample Systems \label{tab:phot}}
\tablehead{\colhead{Filter} & \colhead{Primary Magnitude} & \colhead{Secondary Magnitude} & \colhead{Ref.} \\
& \colhead{(mag)} & \colhead{(mag)} & }
\startdata
\hline
\multicolumn{4}{c}{2MASS J04294155+2632582 (DH Tau)}\\
\hline
F775W       & ...               & $20.2\pm0.03$   & 1 \\
F850LP      & ...               & $18.0\pm0.02$   & 1 \\
$J$         & $9.767\pm0.021$   & $15.71\pm0.05$  & 2,3 \\
$H$         & $8.824\pm0.026$   & $14.96\pm0.04$  & 2,3 \\
$K_s$       & $8.178\pm0.026$   & $14.19\pm0.02$  & 2,3 \\
{[3.6]}     & $7.58\pm0.02$     & $13.32\pm0.24\pm0.56$  & This work \\
{[4.5]}     & $7.21\pm0.02$     & $12.51\pm0.17\pm0.42$  & This work \\
{[5.8]}     & $7.10\pm0.02$     & $11.89\pm0.16\pm0.31$  & This work \\
{[8.0]}     & $6.76\pm0.02$     & $11.13\pm0.15\pm0.19$  & This work \\
\hline
\multicolumn{4}{c}{2MASS J04414565+2301580 (2M0441 AB)}\\
\hline
$y_{\rm P1}$    & $12.06\pm0.01$    & $16.10\pm0.01$ & 4 \\
$J$             & $10.74\pm0.02$    & $14.42\pm0.03$ & 2 \\
$H$             & $10.10\pm0.02$    & $13.73\pm0.03$ & 2 \\
$K_s$           & $9.85\pm0.02$     & $13.16\pm0.03$ & 2 \\
{[3.6]}         & $9.59\pm0.02$     & $12.26\pm0.02\pm0.01$  & This work \\
{[4.5]}         & $9.48\pm0.02$     & $11.89\pm0.02\pm0.01$  & This work \\
{[5.8]}         & $9.32\pm0.02$     & $11.48\pm0.02\pm0.01$  & This work \\
{[8.0]}         & $9.19\pm0.02$     & $10.96\pm0.02\pm0.01$  & This work \\
\hline
\multicolumn{4}{c}{2MASS J06191291--5803156 (AB Pic)}\\
\hline
$B_T$           & $10.24\pm0.02$    & ...           & 5  \\
$G_{\rm RP}$    & $9.29\pm0.01$     & ...           & 6  \\
$G_{\rm BP}$    & $8.21\pm0.01$     & ...           & 6  \\
$J$             & $7.58\pm0.02$     & $16.18\pm0.10$& 2,7 \\
$H$             & $7.09\pm0.02$     & $14.69\pm0.10$& 2,7 \\
$K_s$           & $6.98\pm0.02$     & $14.14\pm0.08$& 2,7 \\
{[3.6]}         & $6.89\pm0.02$     & $13.22\pm0.06\pm0.08$& This work \\
{[4.5]}         & $6.89\pm0.02$     & $12.87\pm0.07\pm0.09$& This work \\
{[5.8]}         & $6.85\pm0.02$     & $12.50\pm0.26\pm0.09$& This work \\
{[8.0]}         & $6.83\pm0.02$     & $12.41\pm0.16\pm0.20$& This work \\
\hline
\multicolumn{4}{c}{2MASS J11062877--7737331 (CHXR 73)}\\
\hline
$G_{\rm RP}$    & $15.76\pm0.01$    & ...               & 6 \\
F625W           & $19.03\pm0.08$    & ...               & 8 \\
F775W           & ...               & $24.07\pm0.13$    & 8 \\
F850LP          & ...               & $21.35\pm0.04$    & 8 \\
$J$             & $12.67\pm0.03$    & $17.87\pm0.30$    & 2,9 \\
$H$             & $11.32\pm0.02$    & $16.52\pm0.30$    & 2,9 \\
$K_s$           & $10.70\pm0.02$    & $15.40\pm0.25$    & 2,9 \\
{[3.6]}         & $10.30\pm0.02$    & $13.94\pm0.05\pm0.12$    & This work \\
{[4.5]}         & $10.24\pm0.02$    & $14.03\pm0.06\pm0.04$    & This work \\
{[5.8]}         & $10.12\pm0.02$    & $>$13.88          & This work \\
{[8.0]}         & $10.14\pm0.02$    & $13.06\pm0.12\pm0.08$    & This work \\
\hline
\multicolumn{4}{c}{2MASS J13164653+0925269 (GJ 504)}\\
\hline
$J$             & $4.13\pm0.02$ & $19.78\pm0.10$   & 10 \\
$H$             & $3.88\pm0.02$ & $20.01\pm0.10$   & 10 \\
$K_s$           & $3.81\pm0.02$ & $19.38\pm0.11$   & 10 \\
$L^{\prime}$    & $3.80\pm0.02$ & $16.70\pm0.17$   & 10  \\
{[3.6]}         & $3.84\pm0.02$ & $>$7.96 & This work \\
{[4.5]}         & $3.90\pm0.02$ & $>$7.35 & This work \\
{[5.8]}         & $3.79\pm0.02$ & $>$8.97 & This work \\
{[8.0]}         & $3.78\pm0.02$ & $>$8.27 & This work \\
\hline
\multicolumn{4}{c}{2MASS J16093030--2104589 (1RXS J1609)}\\
\hline
$I_{\rm DENIS}$             & $10.99\pm0.03$    & ...               & 11   \\
$z^{\prime}_{\rm VISAO}$& $10.60\pm0.06$    & $21.24\pm0.15$    & 12 \\
$Y_{s, \rm VISAO}$      & $10.43\pm0.10$    & ...               & 12 \\
$J$                     & $9.82\pm0.03$     & $17.85\pm0.12$    & 2,13 \\
$H$                     & $9.12\pm0.02$     & $16.86\pm0.07$    & 2,13 \\
$K_s$                   & $8.92\pm0.02$     & $16.15\pm0.05$    & 2,13 \\
$L^{\prime}_{\rm NIRC2}$    & $8.73\pm0.05$           & $14.87\pm0.31$& 14 \\
{[3.6]}                 & $8.77\pm0.02$     & $>$12.88          & This work \\
{[4.5]}                 & $8.79\pm0.02$     & $14.82\pm0.15\pm0.17$    & This work \\
{[5.8]}                 & $8.71\pm0.02$     & $>$13.49          & This work \\
{[8.0]}                 & $8.69\pm0.02$     & $>$14.18          & This work \\
\hline
\multicolumn{4}{c}{2MASS J16271951--2441403 (SR 12)}\\
\hline
$g_{\rm P1}$    & $13.09\pm0.02$	& ...           & 4 \\    
$z_{\rm P1}$    & ...               & $19.03\pm0.04$& 4 \\
$y_{\rm P1}$    & ...               & $17.82\pm0.04$& 4 \\
$J$             & $9.42\pm0.02$     & ...           & 2 \\
$H$             & $8.63\pm0.04$     & ...           & 2 \\
$K_s$             & $8.41\pm0.04$     & $14.42\pm0.07$& 2 \\
{[3.6]}         & $8.16\pm0.02$     & $13.99\pm0.07\pm0.08$& This work \\
{[4.5]}         & $8.10\pm0.02$     & $13.18\pm0.03\pm0.06$& This work \\
{[5.8]}         & $8.05\pm0.02$     & $13.09\pm0.07\pm0.13$& This work \\
{[8.0]}         & $8.01\pm0.02$     & $12.17\pm0.04\pm0.12$& This work \\
\hline
\multicolumn{4}{c}{2MASS J16311501--2432436 (ROXs 42B)}\\
\hline
$g_{\rm P1}$                & $15.07\pm0.01$& ...           & 4 \\
$r_{\rm P1}$                & $13.49\pm0.1$ & ...           & 4 \\
$i_{\rm P1}$                & $12.51\pm0.1$ & ...           & 4 \\
$z_{\rm P1}$                & $11.81\pm0.1$ & ...           & 4 \\
$y_{\rm P1}$                & $11.47\pm0.1$ & ...           & 4 \\
$J$                         & $9.91\pm0.02$ & ...           & 2 \\
$H$                         & $9.02\pm0.02$ & ...           & 2 \\
$K_s$                         & $8.67\pm0.02$ & ...           & 2 \\
$J_{\rm NIRC2}$             & ...           & $16.99\pm0.07$& 14 \\
$H_{\rm NIRC2}$             & ...           & $15.88\pm0.06$& 14 \\
$K^{\prime}_{\rm NIRC2}$           & ...           & $15.01\pm0.05$& 14 \\
$L^{\prime}_{\rm NIRC2}$    & ...           & $14.15\pm0.09$& 14 \\
{[3.6]}                     & $8.38\pm0.02$ & $>$12.29      & This work \\
{[4.5]}                     & $8.36\pm0.02$ & $13.44\pm0.56\pm0.24$& This work \\
{[5.8]}                     & $8.28\pm0.02$ & $>$12.37      & This work \\
{[8.0]}                     & $8.28\pm0.02$ & $12.82\pm0.11\pm0.10$& This work \\
\hline
\multicolumn{4}{c}{2MASS J21185820+2613500 (HD 203030)}\\
\hline
$B_T$       & $9.39\pm0.02$ & ...           & 5  \\
$G_{\rm BP}$& $8.94\pm0.01$ & ...           & 6 \\
$G_{\rm RP}$& $8.69\pm0.01$ & ...           & 6 \\
$J$         & $7.07\pm0.02$ & $18.77\pm0.08$& 2,15 \\
$H$         & $6.73\pm0.02$ & $17.57\pm0.08$& 2,15 \\
$K_s$       & $6.65\pm0.02$ & $16.21\pm0.10$& 2,16 \\
{[3.6]}     & $6.73\pm0.02$ & $15.01\pm0.02\pm0.40$& This work \\
{[4.5]}     & $6.76\pm0.02$ & $14.63\pm0.02\pm0.26$& This work \\
{[5.8]}     & $6.88\pm0.02$ & $13.84\pm0.02\pm0.20$& This work \\
{[8.0]}     & $6.61\pm0.02$ & $13.44\pm0.03\pm0.21$& This work
\enddata
\tablecomments{When two uncertainties are listed, the first is a statistical uncertainty based on the rms between images and the second is a systematic uncertainty based on the PSF-subtraction of the primary. If an entry is missing, that filter was not used in the component's SED fit.}
\tablerefs{(1) \cite{zhou14}; (2) \cite{cutri03}; (3) \cite{itoh05}; (4) \cite{chambers16}; (5) \cite{hog00}; (6) \cite{gaia18}; (7) \cite{chauvin05}; (8) Hubble Legacy Archive; (9) \cite{luhman06c}; (10) \cite{skemer16}; (11) \cite{epchtein97}; (12) \cite{wu15}; (13) \cite{lachapelle15}; (14) \cite{kraus14}; (15) \cite{miles-paez17}; (16) \cite{metchev06}}
\end{deluxetable*}

\begin{deluxetable*}{lcccccccccccccc}
\tablecaption{Companion Contrast Limits}
\tablehead{2MASS & Distance & Ch. & $M$ &Exp.\ Time & \multicolumn{10}{c}{Contrast (mag) at $\rho=$(arcsec)} \\ \cline{6-15}
 & (pc) & & (mag) & (s) & 0.5 & 1.0 & 1.5 & 2.0 & 3.0 & 4.0 & 5.0 & 7.0 & 10.0 & 12.0}
\startdata
J04294155+2632582	& 133.3	& 1	& 1.96	& 0.4	& 4.79	& 4.49	& 4.36	& 4.46	& 5.08	& 5.76	& 6.46	& 6.23	& 6.45	& 6.22	\\
					&		& 2	& 1.66	& 0.4	& 5.02	& 4.75	& 4.62	& 4.55  & 4.85	& 5.76	& 6.00	& 5.59	& 5.73	& 6.03	\\
					&		& 3	& 1.48	& 10.4	& 4.20	& 4.04	& 4.22	& 4.36	& 4.73	& 6.52	& 7.35	& 6.80	& 6.74	& 6.83	\\
					&		& 4	& 1.14	& 10.4	& 4.94	& 4.45	& 4.80	& 5.39	& 4.15	& 4.81	& 5.74	& 6.59	& 6.50	& 6.88	\\
J04414565+2301580	& 122.9	& 1	& 4.13	& 26.8	& 3.19	& 2.15	& 1.87	& 1.64	& 2.43	& 5.72	& 7.90	& 7.53	& 7.63	& 7.59	\\
					&		& 2	& 4.01	& 26.8	& 4.75	& 4.65	& 4.84	& 5.02	& 6.43	& 7.04	& 7.51	& 7.37	& 7.46	& 7.05	\\
					&		& 3	& 3.87	& 26.8	& 6.76	& 5.99	& 6.03	& 6.18	& 6.45	& 7.94	& 8.21	& 7.16	& 7.26	& 6.71	\\
					&		& 4	& 3.74	& 26.8	& 5.37  & 6.08	& 5.52	& 5.47	& 5.26	& 5.54	& 5.90	& 5.89	& 6.56	& 6.37	\\
J06191291--5803156	& 50.1	& 1	& 3.39	& 0.4	& ...	& 6.90	& 6.26	& 6.37	& 6.96	& 7.23	& 7.79	& 7.33	& 7.66	& 7.57	\\
					&		& 2	& 3.39	& 0.4	& 5.89	& 5.81	& 5.99	& 6.19	& 6.57	& 7.60	& 7.01	& 7.28	& 7.18	& 6.99	\\
					&		& 3	& 3.35	& 0.4	& 6.53	& 6.49	& 5.83	& 5.78	& 6.54	& 6.34	& 6.29	& 6.32	& 6.14	& 6.04	\\
					&		& 4	& 3.33	& 0.4	& 5.74	& 5.24	& 5.06	& 4.97	& 5.08	& 5.36	& 5.66	& 6.13	& 6.29	& 6.15	\\
J11062877--7737331	& 190.0	& 1	& 3.91	& 10.4	& 4.73	& 4.55	& 4.45	& 4.60	& 5.19	& 5.88	& 6.65	& 6.84	& 7.01	& 7.34	\\
					&		& 2	& 3.85	& 10.4	& 6.05	& 5.75	& 5.82	& 5.55	& 5.82	& 6.75	& 6.62	& 6.45	& 6.62	& 6.50	\\
					&		& 3	& 3.73	& 10.4	& 3.72	& 3.76	& 3.72	& 3.80	& 3.96	& 4.42	& 5.15	& 5.20	& 5.27	& 5.54	\\
					&		& 4	& 3.75	& 10.4	& 4.29	& 4.14	& 4.15	& 4.10	& 3.72	& 3.88	& 4.97	& 4.43	& 4.54	& 4.32	\\
J13164653+0925269	& 17.6	& 1	& 2.62	& 1.0	& 5.91	& 5.16	& 4.62	& 4.26	& 4.16	& 5.92	& 6.15	& 7.27	& 7.55	& 7.94	\\
					&		& 2	& 2.68	& 1.0	& 5.15	& 4.80	& 4.51	& 3.95	& 3.42	& 4.27	& 5.39	& 7.06	& 7.09	& 7.77	\\
					&		& 3	& 2.58	& 1.0	& 6.71	& 5.18	& 4.97	& 5.01	& 5.39	& 5.40	& 6.85	& 7.33	& 7.18	& 7.28	\\
					&		& 4	& 2.56	& 1.0	& 4.46	& 4.22	& 4.36	& 4.45	& 4.58	& 5.27	& 6.28	& 6.46	& 6.40	& 6.70	\\
J16093030--2104589	& 137.8	& 1	& 3.07	& 10.4	& 4.22	& 3.99	& 3.97	& 4.09	& 4.70	& 6.37	& 7.52	& 7.36	& 7.43	& 7.46	\\
					&		& 2	& 3.09	& 1.2	& 6.16	& 6.18	& 6.29	& 6.40	& 6.45	& 6.94	& 6.97	& 7.12	& 7.08	& 6.93	\\
					&		& 3	& 3.01	& 10.4	& 5.23	& 4.89	& 4.73	& 4.73	& 5.07	& 5.77	& 7.20	& 6.50	& 6.46	& 6.32	\\
					&		& 4	& 2.99	& 1.2	& 5.84	& 5.60	& 5.56	& 5.47	& 5.07	& 5.16	& 5.37	& 5.91	& 5.60	& 5.66	\\
J16271951--2441403	& 112.5	& 1	& 2.90	& 10.4	& 5.87	& 4.91	& 4.55	& 4.55	& 4.90	& 6.13	& 7.47	& 6.96	& 7.06	& 7.02	\\
					&		& 2	& 2.84	& 10.4	& 4.64	& 4.13	& 4.04	& 3.99	& 5.64	& 6.28	& 6.66	& 6.78	& 6.76	& 7.04	\\
					&		& 3	& 2.79	& 10.4	& 4.72	& 4.37	& 4.21	& 4.20	& 4.82	& 5.74	& 5.94	& 5.79	& 5.83	& 6.07	\\
					&		& 4	& 2.75	& 10.4	& 3.32	& 2.70	& 2.33	& 2.29	& 3.53	& 3.26	& 4.01	& 4.14	& 5.35	& 5.36	\\
J16311501--2432436	& 145.4	& 1	& 2.55	& 0.4	& 4.31	& 4.04	& 3.94	& 4.03	& 4.69	& 5.22	& 5.67	& 5.54	& 5.77	& 5.57	\\
					&		& 2	& 2.55	& 0.4	& 5.93	& 5.40	& 5.00	& 4.73	& 4.40	& 4.46	& 4.69	& 4.80	& 5.03	& 5.08	\\
					&		& 3	& 2.47	& 10.4	& 4.39	& 4.11	& 4.03	& 4.15	& 4.94	& 5.60	& 6.89	& 6.62	& 6.97	& 6.78	\\
					&		& 4	& 2.47	& 10.4	& 5.65	& 5.59	& 5.78	& 5.70	& 4.95	& 5.92	& 6.11	& 5.76	& 5.78	& 5.61	\\
J21185820+2613500	& 39.2	& 1	& 3.76	& 26.8	& 7.27	& 4.70	& 4.25	& 4.13	& 4.36	& 4.77	& 6.89	& 7.56	& 7.73	& 7.66	\\
					&		& 2	& 3.79	& 26.8	& ...	& 6.78	& 5.57	& 4.66	& 4.11	& 4.09	& 5.62	& 7.36	& 7.24	& 7.69	\\
					&		& 3	& 3.91	& 26.8	& 1.66	& 1.26	& 1.18	& 1.27	& 1.42	& 2.18	& 3.86	& 6.55	& 7.49	& 7.22	\\
					&		& 4	& 3.64	& 26.8	& 3.28	& 3.05	& 3.05	& 3.15	& 3.38	& 3.90	& 6.85	& 6.40	& 6.66	& 6.98	
\enddata	
\tablecomments{}
\label{tab:contrast_limits}
\end{deluxetable*}

\begin{deluxetable*}{lcccccccccccccc}
\tablecaption{Companion Mass Limits}
\tablehead{2MASS & Distance & Ch. & $M$ &Exp.\ Time & \multicolumn{10}{c}{Mass Limit ($M_{\mathrm{Jup}}$) at $\rho=$(arcsec)} \\ \cline{6-15}
 & (pc) & & (mag) & (s) & 0.5 & 1.0 & 1.5 & 2.0 & 3.0 & 4.0 & 5.0 & 7.0 & 10.0 & 12.0}
\startdata
J04294155+2632582	& 133.3	& 1	& 1.96	& 0.4	& 30	& 35	& 38	& 36	& 25	& 16	& 10	& 12	& 10	& 12	\\
					&		& 2	& 1.66	& 0.4	& 30	& 35	& 37	& 38    & 33	& 19	& 16	& 21	& 19	& 16	\\
					&		& 3	& 1.48	& 10.4	& 47	& 51	& 47	& 44	& 36	& 12	& 6	    & 10	& 11	& 10	\\
					&		& 4	& 1.14	& 10.4	& 37	& 48	& 40	& 29	& 56	& 39	& 23	& 13	& 14	& 11	\\
J04414565+2301580	& 122.9	& 1	& 4.13	& 26.8	& 21	& 38	& 44	& 50	& 33	& $<$1	& $<$1	& $<$1	& $<$1	& $<$1	\\
					&		& 2	& 4.01	& 26.8	& 7	    & 8 	& 6	    & 5 	& $<$1	& $<$1	& $<$1	& $<$1	& $<$1	& $<$1	\\
					&		& 3	& 3.87	& 26.8	& $<$1	& $<$1	& $<$1	& $<$1	& $<$1	& $<$1	& $<$1	& $<$1	& $<$1	& $<$1	\\
					&		& 4	& 3.74	& 26.8	& 2     & $<$1	& 1 	& 2 	& 3 	& 1 	& $<$1	& $<$1	& $<$1	& $<$1	\\
J06191291--5803156	& 50.1	& 1	& 3.39	& 0.4	& ...	& 9 	& 11	& 11	& 8	    & 7 	& 5	    & 7	    & 5	    & 6	\\
					&		& 2	& 3.39	& 0.4	& 13	& 13	& 13	& 12	& 10	& 7 	& 9	    & 8	    & 8	    & 9	\\
					&		& 3	& 3.35	& 0.4	& 10	& 10	& 13	& 13	& 10	& 11	& 11	& 11	& 11	& 12	\\
					&		& 4	& 3.33	& 0.4	& 12	& 15	& 15	& 15	& 15	& 14	& 13	& 11	& 10	& 10	\\
J11062877--7737331	& 190.0	& 1	& 3.91	& 10.4	& 9 	& 10	& 11	& 10	& 5	    & $<$1	& $<$1	& $<$1	& $<$1	& $<$1	\\
					&		& 2	& 3.85	& 10.4	& $<$1	& 1	    & 1	    & 3	    & 1	    & $<$1	& $<$1	& $<$1	& $<$1	& $<$1	\\
					&		& 3	& 3.73	& 10.4	& 17	& 17	& 17	& 16	& 14	& 11	& 6	    & 5	    & 5 	& 3	\\
					&		& 4	& 3.75	& 10.4	& 11	& 12	& 12	& 12	& 16	& 14	& 6	    & 10	& 9 	& 11	\\
J13164653+0925269	& 17.6	& 1	& 2.62	& 1.0	& 140	& 190	& 240	& 280	& 300	& 140	& 120	& 82	& 78	& 75	\\
					&		& 2	& 2.68	& 1.0	& 180	& 200	& 230	& 300	& 400	& 260	& 160	& 84	& 84	& 76	\\
					&		& 3	& 2.58	& 1.0	& 91	& 180	& 190	& 190	& 160	& 160	& 88	& 80	& 83	& 81	\\
					&		& 4	& 2.56	& 1.0	& 240	& 270	& 260	& 250	& 230	& 170	& 100	& 96	& 98	& 89	\\
J16093030--2104589	& 137.8	& 1	& 3.07	& 10.4	& 39	& 49	& 50    & 45	& 22	& 12    & 7	    & 8	    & 8	    & 8	\\
					&		& 2	& 3.09	& 1.2	& 13	& 13	& 13    & 12	& 12	& 10	& 10	& 10    & 10 	& 10	\\
					&		& 3	& 3.01	& 10.4	& 17	& 19	& 20    & 20	& 18	& 14    & 9 	& 11	& 12	& 12	\\
					&		& 4	& 2.99	& 1.2	& 13	& 14	& 15    & 15	& 17	& 17	& 16	& 13	& 14	& 14	\\
J16271951--2441403	& 112.5	& 1	& 2.90	& 10.4	& 10	& 19	& 23	& 23	& 19	& 8	    & $<$1	& 2     & 1	    & 1	\\
					&		& 2	& 2.84	& 10.4	& 21	& 26	& 27	& 28	& 12	& 7     & 4	    & 4	    & 4	    & 2	\\
					&		& 3	& 2.79	& 10.4	& 19	& 23	& 25	& 25	& 18	& 11	& 9	    & 10	& 10	& 8	\\
					&		& 4	& 2.75	& 10.4	& 38	& 53	& 68	& 69	& 34	& 39	& 27	& 26	& 13	& 13	\\
J16311501--2432436	& 145.4	& 1	& 2.55	& 0.4	& 30	& 33	& 35	& 34	& 25	& 19	& 14	& 15	& 14	& 15	\\
					&		& 2	& 2.55	& 0.4	& 12	& 16	& 20	& 23	& 27	& 26	& 23	& 22	& 20	& 19	\\
					&		& 3	& 2.47	& 10.4	& 27	& 30	& 32	& 30	& 21	& 14	& 4     & 6     & 4     & 5     \\
					&		& 4	& 2.47	& 10.4	& 13	& 13	& 12	& 12	& 19	& 11	& 9     & 12	& 12	& 13	\\
J21185820+2613500	& 39.2	& 1	& 3.76	& 26.8	& 18	& 72	& 100	& 110	& 91	& 69	& 22	& 15	& 13	& 14	\\
					&		& 2	& 3.79	& 26.8	& ...	& 24	& 40	& 69	& 100	& 100	& 39	& 17	& 19	& 13	\\
					&		& 3	& 3.91	& 26.8	& 460	& 550	& 560	& 540	& 520	& 350	& 110	& 23	& 8     & 13	\\
					&		& 4	& 3.64	& 26.8	& 190	& 230	& 230	& 210	& 180	& 130	& 18	& 24	& 21	& 15	
\enddata
\tablecomments{}
\label{tab:mass_limits}
\end{deluxetable*}

\begin{deluxetable*}{lcc}
\tablecaption{SED-fitting Results for Sample Systems}
\tablehead{\colhead{Parameter} & \colhead{Primary} & \colhead{Companion}}
\startdata
\hline
\multicolumn{3}{c}{2MASS J04294155+2632582 (DH Tau)}\\
\hline
$T_{\mathrm{eff}}$ (K)  & $2600\pm100$  & $1800\pm50$  \\
$E(B-V)$ (mag)                & $1.27\pm0.06$& $0.00\pm0.03$  \\
$\chi^2_{\nu}$          & 0.65          & 6.03  \\
${[8.0]}_{mod}-{[8.0]}_{obs}$ (mag)& 0.41    & 1.94  \\
\hline
\multicolumn{3}{c}{2MASS J04414565+2301580 (2M0441)}\\
\hline
$T_{\mathrm{eff}}$ (K)  & $3200\pm100$  & $2800\pm50$  \\
$E(B-V)$ (mag)                & $0.00\pm0.02$ & $0.46\pm0.04$ \\
$\chi^2_{\nu}$          & 0.36          & 3.11  \\
${[8.0]}_{mod}-{[8.0]}_{obs}$ (mag)& 0.22     & 1.50  \\
\hline
\multicolumn{3}{c}{2MASS J06191291--5803156 (AB Pic)}\\
\hline
$T_{\mathrm{eff}}$ (K)  & $6000\pm100$  & $2100\pm100$  \\
$E(B-V)$ (mag)                & $0.30\pm0.02$ & $1.70\pm0.19$  \\
$\chi^2_{\nu}$          & 4.12          & 0.84  \\
${[8.0]}_{mod}-{[8.0]}_{obs}$ (mag)& 0.08    & 0.21  \\
\hline
\multicolumn{3}{c}{2MASS J11062877--7737331 (CHXR 73)}\\
\hline
$T_{\mathrm{eff}}$ (K)  & $3700\pm50$  & $1700\pm50$    \\
$E(B-V)$ (mag)                & $2.12\pm0.03$& $1.77\pm0.11$  \\
$\chi^2_{\nu}$          & 1.69  & 0.45  \\
${[8.0]}_{mod}-{[8.0]}_{obs}$ (mag)& $-0.07$    & $0.01$  \\
\hline
\multicolumn{3}{c}{2MASS J13164653+0925269 (GJ 504)}\\
\hline
$T_{\mathrm{eff}}$ (K)  & $6900\pm200$  & $800\pm100$  \\
$E(B-V)$ (mag)                & $0.16\pm0.04$ & $0.00\pm0.18$   \\
$\chi^2_{\nu}$          & 2.78          & 12.17  \\
${[8.0]}_{mod}-{[8.0]}_{obs}$ (mag)& 0.06    & $<$8.46  \\
\hline
\multicolumn{3}{c}{2MASS J16093030--2104589 (1RXS J1609)}\\
\hline
$T_{\mathrm{eff}}$ (K)  & $3700\pm50$ & $2000\pm100$  \\
$E(B-V)$ (mag)                & $0.00\pm0.01$ & $2.30\pm0.13$   \\
$\chi^2_{\nu}$          & 0.60      & 3.92  \\
${[8.0]}_{mod}-{[8.0]}_{obs}$ (mag)& 0.02    & $<$0.24  \\
\hline
\multicolumn{3}{c}{2MASS J16271951--2441403 (SR 12)}\\
\hline
$T_{\mathrm{eff}}$ (K)  & $3600\pm50$   & $2200\pm50$   \\
$E(B-V)$ (mag)                & $0.22\pm0.01$ & $0.14\pm0.03$ \\
$\chi^2_{\nu}$          & 1.14          & 4.24          \\
${[8.0]}_{mod}-{[8.0]}_{obs}$ (mag)& 0.04    & 1.35  \\
\hline
\multicolumn{3}{c}{2MASS J16311501--2432436 (ROXs 42B)}\\
\hline
$T_{\mathrm{eff}}$ (K)  & $3600\pm50$   & $1700\pm50$  \\
$E(B-V)$ (mag)                & $0.50\pm0.02$ & $0.49\pm0.18$  \\
$\chi^2_{\nu}$          & 4.29          & 4.20  \\
${[8.0]}_{mod}-{[8.0]}_{obs}$ (mag)& 0.01    & 0.46  \\
\hline
\multicolumn{3}{c}{2MASS J21185820+2613500 (HD 203030)}\\
\hline
$T_{\mathrm{eff}}$ (K)  & $5600\pm50$   & $1700\pm50$   \\
$E(B-V)$ (mag)                & $0.01\pm0.02$ & $2.61\pm0.30$ \\
$\chi^2_{\nu}$          & 8.79          & 0.68  \\
${[8.0]}_{mod}-{[8.0]}_{obs}$ (mag)& 0.13    & 0.56  
\enddata
\tablecomments{}
\label{tab:sedfit}
\end{deluxetable*}

\end{document}